\documentclass[showpacs,aps,prd,floatfix,amsmath,amssymb]{revtex4}
\usepackage{graphicx}
\usepackage{epstopdf}
\usepackage[utf8]{inputenc}
\usepackage{amsmath}
\usepackage{array}
\usepackage{slashed}	
\usepackage[makeroom]{cancel}
\begin{document}
\title{Decays  $A \to Z\gamma\gamma$ and $\phi \to Z\gamma\gamma$ ($\phi=h,H$) in two-Higgs doublet models}
\author{R. S\'anchez-V\'elez}
\author{G. Tavares-Velasco}
\email[Corresponding author: ]{gtv@fcfm.buap.mx}
\affiliation{Facultad de
Ciencias F\'\i sico Matem\'aticas, Benem\'erita Universidad
Aut\'onoma de Puebla, Apartado Postal 1152, Puebla, Pue., M\'exico}

\begin{abstract}
The one-loop contributions to the decays of the $CP$-odd and  $CP$-even scalar bosons $A\to Z\gamma\gamma$ and $\phi\to Z\gamma\gamma$ ($\phi=h,H$) are calculated  within the framework of $CP$-conserving THDMs, where they are induced by  box and reducible Feynman diagrams.  The behavior of the corresponding branching ratios are then analyzed within the type-II THDM in a region of the parameter space around the alignment limit and still consistent with experimental data.  It is found that the $A\to Z\gamma\gamma$ branching ratio is only relevant when $m_A>m_H+m_Z$, but it is negligible otherwise.  For $m_A>600$ GeV and $t_\beta\simeq O(1)$, $BR(A\to Z\gamma\gamma)$ can reach values of the order of $10^{-5}-10^{-4}$, but it decreases by about one order of magnitude as $t_\beta$ increases up to 10.  A similar behavior is followed by the $H\to Z\gamma\gamma$ decay, which only has a non-negligible branching ratio when  $m_H>m_A+m_Z$ and can reach the level of $10^{-4}-10^{-3}$ for $m_H>600$ GeV and $t_\beta\simeq O(1)$.  We also estimated the branching ratios of these rare decays in the type-I THDM, where they can be about one order of magnitude larger than in type-II THDM. As far as the $h\to Z\gamma\gamma$ decay is concerned, since the properties of this scalar boson must be nearly identical to those of the SM Higgs boson, the $h\to Z\gamma\gamma$ branching ratio does not deviates significantly from the SM prediction, where it is negligibly small, of the order of $10^{-9}$. This result is in agreement with previous calculations.

\end{abstract}
\maketitle

\section{Introduction}

The standard model (SM) has provided  a successful description of the observed electroweak phenomena at the energy scales explored until now, as confirmed recently with the discovery of the Higgs boson by  the ATLAS and CMS experiments at the CERN LHC \cite{Aad:2012tfa,Chatrchyan:2012xdj}. Nonetheless, it is worth to explore whether there is a unique Higgs boson, as predicted by the SM, or  the electroweak symmetry breaking  (EWSB) mechanism requires additional Higgs bosons. To address some SM flaws, a plethora of extension models  have been proposed, several of which contain a scalar sector with more than one Higgs multiplet, thereby predicting more than one physical Higgs boson. If experimental data reveal the existence of any additional Higgs bosons, it will be crucial to test what extension model is consistent with such  particles. The simplest of such theories are  two-Higgs doublet models (THDMs)  \cite{Gunion:1989we,Branco:2011iw}, which are obtained by adding a second complex $SU(2)_L$ Higgs doublet to the  SM one. These  models respect the $\rho=1$ relation  at the tree-level, contrary to other higher-dimensional Higgs-multiplet models. Also, in spite of its simplicity, THDMs can  predict several new phenomena absent in the SM, such as new sources of $CP$ violation, tree-level scalar-mediated flavor changing neutral currents (FCNCs), a dark matter candidate, etc.  After  EWSB, three of the eight degrees of freedom are removed from the spectrum to provide the longitudinal modes of the  $W^\pm$ and $Z$ gauge bosons. Five physical Higgs bosons remain as remnant: a charged Higgs boson pair $H^\pm$ and three neutral Higgs bosons $h$, $H$, and $A$. If the scalar sector respects $CP$  invariance, the neutral  scalar bosons are $CP$-eigenstates: $h$ and $H$ are $CP$-even, whereas $A$  is $CP$-odd. It is usually assumed that one of the neutral $CP$-even scalar bosons is the one observed at the LHC. The most general $CP$-conserving THDMs have tree-level FCNCs \cite{Atwood:1996vj}, which can be removed by imposing  a $Z_2$ discrete symmetry
that forbids such interactions at the tree-level \cite{PhysRevD.15.1958}. In this scenario, there are four  THDM types, which are typically known as type-I, type-II, lepton-specific \cite{Cao:2009as} and flipped  THDM \cite{Aoki:2009ha}. It turns out that  type-II THDM is  the most studied in the literature as it has the same Yukawa couplings as the minimal supersymmetric standard model (MSSM), therefore  its still-allowed region of  parameter space  has been considerably studied.

Since the proposal of the Higgs mechanism, the phenomenology of the  Higgs bosons has been the focus of considerable attention. As for the dominant   tree-level decay modes of  a $CP$-even  Higgs boson $h\to \bar ff$ and $h\to VV$ ($V=W,Z$), they have been long studied in the literature both in the SM and  several of its extensions, along with the one-loop induced decays $h\to \gamma\gamma$, $h\to \gamma Z$, and $h\to gg$. Although the   $h\to \gamma\gamma$ decay  has a  tiny branching ratio for a $125$ GeV Higgs boson, it was very helpful for the detection of the SM Higgs boson. This decay mode has the advantage of a relatively low background, so it  was fundamental in the design of the ATLAS and CMS detectors. As for the $h\to gg$ decay, it is undetectable but it is fundamental to compute the cross section for  Higgs production via gluon fusion.

It is expected that the data collected at the LHC may allow us to search for any other rare decays of the Higgs boson \cite{Kling:2016opi}, such as  lepton flavor changing Higgs decays $h\to \bar\ell_i \ell_j$ ($i\ne j$) or invisible Higgs decays $h\to \cancel{E}_T$, which are forbidden in the SM and can shed light on any new physics effect.  Even more, with the prospect of a future Higgs boson factory, other exotic decays of the Higgs boson could be at the reach of experimental detection. In particular,   the rare decay $h\to Z\gamma\gamma$   is very suppressed in the SM as it arises at the one-loop level via the exchange of charged particles, so it can offer a relatively clean signal of new physics: two energetic photons plus a back to back lepton anti-lepton pair. This process  can also  provide a test for the  couplings of the Higgs boson to the particles running into the loops, which can be SM particles or any new charged particle predicted by other extension models. A similar decay is $h\to Z gg$, which at the leading order can be straightforwardly calculated from the $h\to Z\gamma\gamma$ one. In addition, the study of the  $hZ\gamma\gamma$ and  $hZ gg$ vertices would allow us to obtain the leading order contributions to the cross section of   $hZ$ pair production via photon fusion   $\gamma\gamma\to hZ$   and gluon fusion  $gg\to hZ$  \cite{Kniehl:1990iva,Gounaris:2001rk}.

On the other hand,  a $CP$-odd scalar boson has  fewer  decay channels and so it is worth studying some one-loop induced decays of such a particle. At tree level, its dominant decay channels are  $A\to \bar f f$, $A\to Zh (H)$ and $A\to W^\pm H^\mp$,  when kinematically allowed, whereas at the one-loop level a $CP$-odd scalar boson can decay as $A\to gg$, $A\to \gamma\gamma$ and $A\to Z\gamma$ \cite{Gunion:1989we}. These decay channels can have  significant branching ratios in  some regions of the parameter space. Other one-loop induced decay modes such as $A\to WW$ and $A\to ZZ$ have already been studied in \cite{Mendez:1991gp,Diaz-Cruz:2014aga}, though they are more suppressed than the aforementioned decay channels.

In this work we are interested in studying the  $A \to Z\gamma\gamma$ and $\phi \to Z\gamma\gamma$ ($\phi=h,H$)  decay modes  in the context of THDMs,  which induce these processes at the one-loop level via   box and reducible Feynman diagrams, with contributions from charged fermions, mainly from the top and bottom quarks. The $W$ gauge boson and the charged scalar boson $H^\pm$ can  only contribute through reducible diagrams to the $A \to Z\gamma\gamma$  decay . The  respective decay of  the SM Higgs boson has already been studied: the decay $h\to Z\gamma\gamma$  was studied in Ref. \cite{Abbasabadi:2004wq} and the analogue decay $h\to Zgg$ was studied in \cite{Kniehl:1990yb,Abbasabadi:2008zz}. To our knowledge, the  $A \to Z\gamma\gamma$  decay has not been studied until now.

The organization of this paper is as follows. Section II is devoted to a brief discussion of the general THDM, focusing on the $CP$-conserving   THDMs. In Sec. III we present the details of the calculation of the decays  $A \to Z\gamma\gamma$ and $\phi \to Z\gamma\gamma$ ($\phi=h,H$) by the Passarino-Veltman  reduction scheme. We present the analytical expressions for the invariant amplitudes, the decay widths, as well as the kinematic distributions of the invariant mass of the photons and the energy of the $Z$ gauge boson, which can be useful to disentangle the decay signal from its potential background. The numerical analysis of the branching ratios within type-II THDM is presented in Sec. IV, whereas    the conclusions and outlook are presented in Sec. V.  The Feynman rules necessary for the calculations and some lengthy formulas are presented in the appendices.

\section{Two-Higgs doublet models}

THDMs have been largely studied in the literature  \cite{Gunion:1989we}. We will present here a brief outline of $CP$-conserving THDMs, including only those details relevant for our calculation. For the interested reader, a comprehensive review of these models can be found in \cite{Branco:2011iw}.

\subsection{THDM Lagrangian}

In THDMs,  two complex $SU(2)_L$ Higgs doublets $\Phi_i$ are introduced in the scalar sector:

\begin{equation}
\Phi_i=\left (\begin{array}{c}
\phi_i^+\\
{\dfrac{v_i+\phi_i^0+i\phi_i}{\sqrt{2}}}
\end{array}\right )\;\;\; (i=1,2),
\end{equation}
where $v_i$ are the vacuum expectation values (VEVs) of the neutral components, which satisfy $v_1^2+v_2^2=v^2$, with $v=246$ GeV. A well known parameter of this model  is  the VEVs ratio $\tan\beta\equiv t_\beta=v_2/v_1$. The  EWSB mechanism is achieved by the most general $SU(2)_L\times U(1)_Y$ gauge invariant Lagrangian

\begin{equation}\label{Lag}
\mathcal{L}=\sum_i |D_\mu\Phi_i|^2-V(\Phi_1,\Phi_2)+\mathcal{L}_{Yuk}+\mathcal{L}_{SM},
\end{equation}
where $|D_\mu\Phi_i|^2$ is the kinetic term for the two-Higgs doublets, with $D_\mu$
the SM covariant derivative,   $V(\Phi_1,\Phi_2)$ is the Higgs potential, $\mathcal{L}_{Yuk}$ denotes the Yukawa interactions between $\Phi_i$ and the SM fermions, and $\mathcal{L}_{SM}$ describes the $SU(2)_L\times U(1)_Y$ interactions of fermions and gauge bosons.

The most general gauge-invariant renormalizable potential $V(\Phi_1,\Phi_2)$ for THDMs is a  hermitian  combination of electroweak invariant combinations. It contains 14 parameters and can give  rise to new sources of $CP$ violation \cite{Ginzburg:2004vp}. However, as  long as $CP$  is conserved in the Higgs sector, the scalar potential for the two doublets $\Phi_1$ and $\Phi_2$ with hypercharge $+1$ can be written in terms of 8 parameters as follows \cite{Gunion:1989we,Branco:2011iw}

\begin{equation}\label{pot}
\begin{split}
V(\Phi_1,\Phi_2)&=m_{11}^2\Phi_1^\dagger\Phi_1+m_{22}^2\Phi_2^\dagger\Phi_2-m_{12}^2(\Phi_1^\dagger\Phi_2+\mbox{h.c.})+\dfrac{\lambda_1}{2}(\Phi_1^\dagger\Phi_1)^2+\dfrac{\lambda_2}{2}(\Phi_2^\dagger\Phi_2)^2\\
&+\lambda_3(\Phi_1^\dagger\Phi_1)(\Phi_2^\dagger\Phi_2)+\lambda_4(\Phi_1^\dagger\Phi_2)(\Phi_2^\dagger\Phi_1)+\dfrac{\lambda_5	}{2}\Bigl[(\Phi_1^\dagger\Phi_2)^2+\mbox{h.c.}\Bigr]
\end{split}
\end{equation}
After EWSB, three of the eight degrees of freedom of the two Higgs doublets are the Goldstone bosons $(G^\pm,\xi)$, which are absorbed as longitudinal components of the $W^\pm$ and $Z$ gauge bosons, whereas the remaining  five degrees of freedom become the physical Higgs bosons: there is a pair of charged scalar bosons $H^\pm$, two neutral $CP$-even scalar bosons $h$ and $H$, where $m_h<m_H$  by convention, and one neutral $CP$-odd scalar $A$. Since all the parameters appearing in the potential are real, there are no bilinear mixing terms, which is why  the neutral mass eigenstates are also  $CP$ eigenstates. In the neutral sector the following mass term appears

\begin{equation}
\mathcal{L}_{\mbox{mass}}^{A}=(\phi_1\;,\; \phi_2)V_A^2\left (\begin{array}{c}
\phi_1\\
\phi_2
\end{array}\right ),
\end{equation}
with
\begin{equation}
V_A^2=\dfrac{1}{2}\left (\frac{m_{12}^2}{v_1 v_2}-\lambda_5\right )\left (\begin{array}{cc}
\;\;v_2^2	&-v_1v_2\\
-v_1v_2	&\;\;v_1^2
\end{array}\right ).
\end{equation}
Once $V_A^2$ is diagonalized, one obtains the neutral Goldstone boson $\xi$ and the physical $CP$-odd Higgs boson via the rotation
\begin{equation}
\left (\begin{array}{c}
\xi\\
A
\end{array}\right )=\left (\begin{array}{rc}
\cos\beta	&\sin\beta\\
-\sin\beta	&\cos\beta
\end{array}\right )\left (\begin{array}{c}
\phi_1\\
\phi_2
\end{array}\right ),
\end{equation}
with $
m_A^2=\left (\dfrac{m_{12}^2}{v_1 v_2}-\lambda_5\right )v^2$.

In the case of the $CP$-even scalar bosons we have

\begin{equation}
\mathcal{L}_{\mbox{mass}}^{H}=\dfrac{1}{2}(\phi_1^0\;\; \phi_2^0)V_H^2\left (\begin{array}{c}
\phi_1^0\\
{\phi_2^0}
\end{array}\right ),
\end{equation}
where
\begin{equation}
V_H^2=\left (\begin{array}{cc}
A_s	&B_s\\
B_s	&C_s
\end{array}\right ),
\end{equation}
with
$A_s=\lambda_1v_1^2+\dfrac{v_2}{v_1}m_{12}^2$,
$B_s=m_{12}^2-\dfrac{v_1}{v_2}(v_1^2\lambda_1-2m_{11}^2)$, and
$C_s=\lambda_2v_2^2+\dfrac{v_1}{v_2}m_{12}^2$. The physical $CP$-even Higgs bosons with masses $m_H$ y $m_h$ are obtained by rotating the original basis by an angle $\alpha$

\begin{equation}
\left (\begin{array}{c}
H\\
h
\end{array}\right )=\left (\begin{array}{rc}
\cos\alpha		&\sin\alpha\\
-\sin\alpha		&\cos\alpha
\end{array}\right ) \left (\begin{array}{c}
\phi_1^0\\
{\phi_2^0}
\end{array}\right ),
\end{equation}
with $m_{H,h}^2=\dfrac{1}{2}\left ((A_s+C_s)\pm\sqrt{(A_s-C_s)^2+B_s^2}\right )$ and the mixing angle given as

\begin{equation}
\label{alpha}
\sin2\alpha=\frac{2B_s}{\sqrt{(A_s-C_s)^2+4B_s^2}}.
\end{equation}

\subsection{Flavor-conserving THDMs}

As far as the Yukawa Lagrangian $\mathcal{L}_{Yuk}$ is concerned, the scalar-to-fermion couplings are not univocally determined by the gauge structure of the model. The most general Yukawa Lagrangian for THDMs is \cite{Branco:2011iw}

\begin{equation}\label{Lyuk}
-\mathcal{L}_{Yuk}=\sum_{k=1}^2\left[\bar{L}_L\Phi_k Y^\ell_k \ell_R+\bar{Q}_L\left( \Phi_k Y^d_k d_R+\tilde{\Phi}_k Y^u_k u_R\right)\right]
+{\rm H.c.},
\end{equation}
where $\tilde \Phi_j=i\tau_2\Phi_j$, $Y^f$ are $3\times 3$ complex matrices, and the left- and right-handed fermion fields are tree-vectors in flavor space.

To prevent tree-level FCNCs it is usual to introduce a discrete $Z_2$ symmetry respected by the $\Phi_i$  doublets and  the fermions. Under this symmetry one of the scalar doublets is even $\Phi_2\rightarrow \Phi_2$ and the other one is odd $\Phi_1\rightarrow -\Phi_1$. This gives rise to four types of THDMs, which are usually known as  type-I THDM, type-II THDM, lepton-specific THDM and flipped THDM. The way in which each Higgs doublet couples to the fermions in these models is summarized in Table \ref{tabyuk}. On the other hand, if no $Z_2$ discrete symmetry is imposed, there will be tree-level FCNCs. In such a scenario both doublets couple to the charged leptons and quarks. This model is known as type-III THDM \cite{Cheng:1987rs,1742-6596-259-1-012073}. In this work, however, we are not interested in this realization of THDMs.

\begin{table}[!hbt]
\caption{Couplings of quarks and leptons to the Higgs doublets $\Phi_i$ in THDMs with natural flavor conservation. The superscript $i$ stands for the generation index. There is another version of THDMs, known as type-III THDM,  in which both Higgs doublets couple to the leptons and quarks simultaneously, thereby giving rise to tree-level FCNCs \cite{Gunion:1989we,Branco:2011iw}. }
\label{tabyuk}
\begin{tabular}[t]{c ccc}
\hline
\hline
THDM	&$u^i$		&$d^i$		&$e^i$\\
\hline
\hline
type-I 	&$\Phi_2$	&$\Phi_2$	&$\Phi_2$\\
\hline
type-II &$\Phi_2$	&$\Phi_1$	&$\Phi_1$\\
\hline
lepton-specific	&$\Phi_2$	&$\Phi_2$	&$\Phi_1$\\
\hline
flipped	&$\Phi_2$	&$\Phi_1$	&$\Phi_2$\\
\hline
\end{tabular}
\end{table}

Although we will present a rather general calculation within flavor-conserving THDMs, the numerical analysis will be carried out in the context of the type-II THDM, which is by far the most studied THDM since it shares the same Yukawa interactions as the MSSM. The most distinctive difference between the type-II THDM and the MSSM is that the former does not have a strict upper bound on the mass of the lightest Higgs boson, which is an important feature of the latter. In addition, in THDMs the scalar  boson self couplings are arbitrary and so is  the mixing parameter $\alpha$, which in the MSSM is given in terms of $\tan\beta$ and the scalar boson masses.

Our calculation is to be performed in the unitary gauge.  The  Feynman rules  for THDMs can be obtained once the Lagrangian is expanded in terms of mass eigenstantes and  can be found for instance in Refs. \cite{Branco:2011iw,Gunion:1989we}.  We present those  Feynman rules required by our calculation in Appendix \ref{FeynmanRules}.

\section{ $A \to Z\gamma\gamma$ and $\phi \to Z\gamma\gamma$ ($\phi=h,H,$) decay widths}
\subsection{Kinematic conditions}

We now turn to present the  $A \to Z\gamma\gamma$ and $\phi \to Z\gamma\gamma$ ($\phi=h,H,$) decay widths. We first present the kinematics conditions, which are defined according to the following  notation for the  external 4-momenta

\begin{equation}
\phi(p)\to \gamma_\mu(k_1)+\gamma_\nu(k_2)+Z_\alpha(k_3).
\end{equation}
The mass-shell conditions thus read $p^2=m_\phi^2$, $k_3^2=m_Z^2$ and $k_1^2=k_2^2=0$. We now introduce the following Lorentz invariant quantities:

\begin{align}
\label{s1}
s_1& = (k_1+k_3)^2,\\
\label{s2}
s_2 & =(k_2+k_3)^2,\\
\label{s}
s &= (k_1+k_2)^2.
\end{align}
These variables are not all independent as $s_1+s_2+s=m_\phi^2+m_Z^2$ by four-momentum conservation.
In our calculation, we express all the scalar products between the four-momenta $k_1$, $k_2$ and $k_3$ in terms of the Lorentz invariant variables $s_1$, $s_2$ and  $s$ as well as the scaled variable $\mu_Z=m_Z^2/m_\phi^2$.

In addition, because of the transversality conditions obeyed by the gauge bosons, i.e., $k_1\cdot \epsilon^\mu(k_1)=k_2\cdot \epsilon^\nu(k_2)=k_3\cdot \epsilon^\alpha(k_3)=0$, we  drop from the invariant amplitudes any terms proportional to $k_1^\mu$, $k_2^\nu$, and $k_3^\alpha$.

All the above kinematic conditions  probe useful to simplify the calculation. We  now present the invariant amplitudes for the $A \to Z\gamma\gamma$ and $\phi \to Z\gamma\gamma$ ($\phi=h,H$) decays, which are induced at the one-loop level at the lowest order in perturbation theory.

\subsection{ $A \to Z\gamma\gamma$ decay invariant amplitude}

There are  two sets of  Feynman diagrams that induce this decay: box diagrams and reducible diagrams. Once the invariant amplitude for each Feynman diagram was written down in the unitary gauge, we used the Passarino-Veltman reduction scheme   to solve the loop integrals \cite{Passarino:1978jh}, which were reduced  down to a combination of two-, three- and four-point scalar functions. The algebra was carried out  with the aid of the Mathematica package  FeynCalc \cite{Mertig:1990an}. We   first present the invariant amplitude arising from the box diagrams.

 \subsubsection{Box diagram contribution}

\begin{figure}[ht!]
 \centering
 \includegraphics[width=12cm]{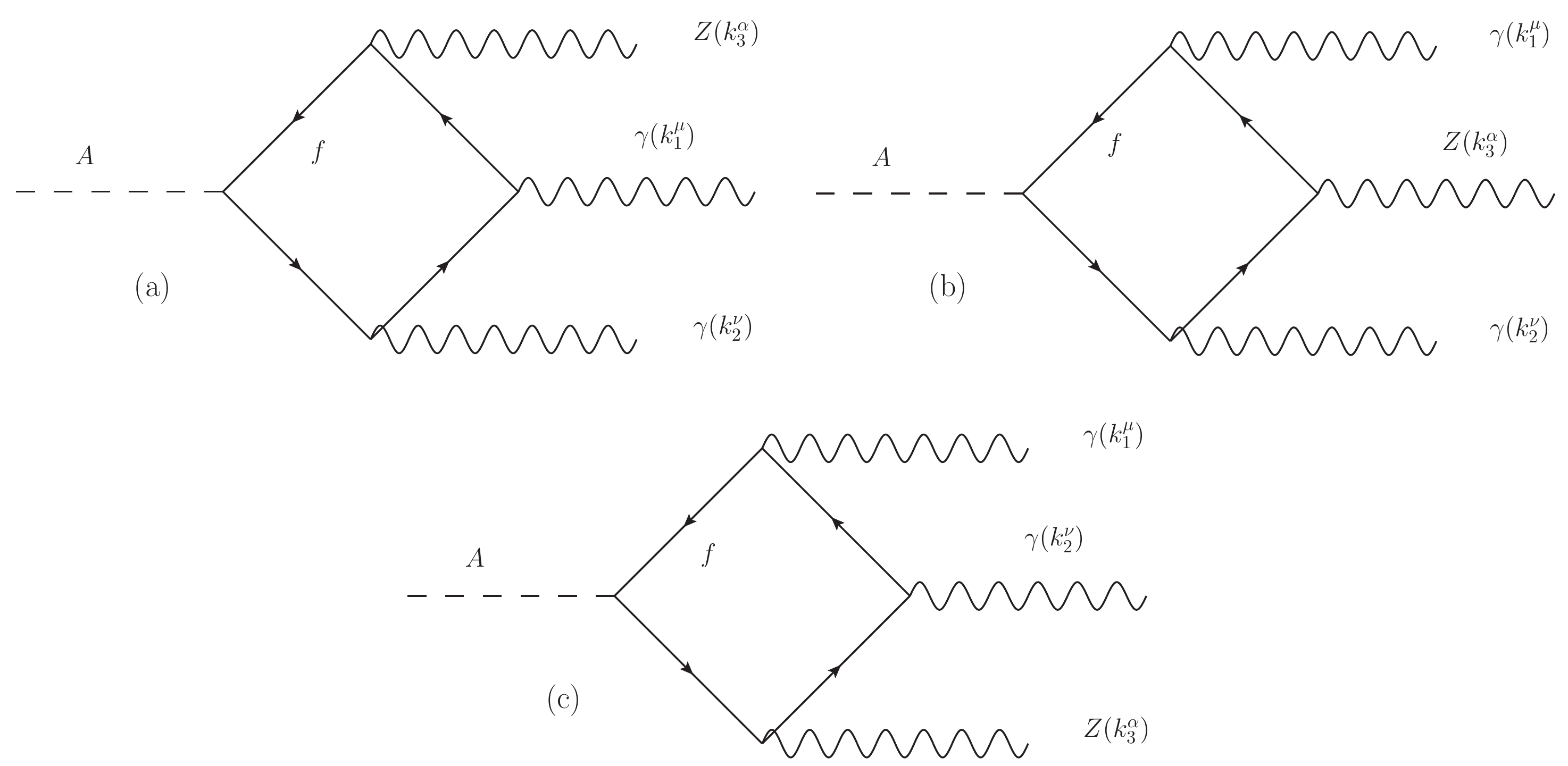}
 \caption{Box diagrams that contribute to the  $A\to Z \gamma\gamma$ decay in the THDM. There are three additional diagrams that are obtained by  exchanging the photons. Similar diagrams also contribute to the $\phi \to Z\gamma\gamma$ ($\phi=h,H$) decay, after the replacement  $A \to \phi$.}
 \label{FeynmanDiagramsHtoZggBox}
\end{figure}

In Fig. \ref{FeynmanDiagramsHtoZggBox} we show the  box diagrams  that contribute to the $A \to Z    \gamma\gamma$  decay.  The dynamical content  is rather simple in the sense that there is only one kind of particles circulating into the loop, namely, SM charged fermions. Other charged particles do not contribute to this decay at the one-loop level in THDMs: due to $CP$ invariance in the scalar sector, the $CP$-odd scalar $A$ does not couple to a pair of $W$ gauge bosons or charged scalars $H^\mp$, though it can couple to a $W^\pm H^\mp$ pair. However, the $V W^\mp H^\pm$ vertex ($V=\gamma, Z$) is absent at the tree-level and so the $A\to Z\gamma\gamma$ decay cannot proceed via box diagrams with  both $W^\mp$ and $H^\mp$ particles.    The main contributions of box diagrams are thus expected to arise from the heaviest fermions. For small $t_\beta$, the  top quark contribution would dominate, whereas for large  $t_\beta$ the  bottom quark contribution would  become relevant. This is due to the presence of the factors $1/t_\beta$ and $ t_\beta$ appearing  in the Yukawa couplings for the top and bottom quarks, respectively, as will be shown below.

Once the Passarino-Veltman reduction scheme was applied, we performed several test on our results. First, we  verified that the  invariant amplitude for all the box diagrams is  gauge invariant under $U(1)_{\mbox{em}}$, i.e. it vanishes when the photon four-momenta are replaced by their polarization vectors. We also verified that   Bose symmetry is respected and that ultraviolet divergences cancel out. The invariant amplitude for the $A\to Z \gamma \gamma$ decay can  be cast in   the following  gauge-invariant manifest form

\begin{equation}\label{AMH}
\mathcal{M}(A\to Z \gamma \gamma)=\mathcal{M}^{\alpha\mu\nu}(A\to Z \gamma \gamma) \;\epsilon_{\alpha}^*(k_3)\epsilon_{\mu}^*(k_1)\epsilon_{\nu}^*(k_2) ,
\end{equation}
with the Lorentz structures given  as follows
\begin{equation}
\begin{split}
\label{AtoZggAmplitude}
\mathcal{M}^{\alpha\mu\nu}(A\to Z \gamma \gamma)&=\mathcal{F}_{1} \;k_1^\alpha\Bigl(k_1^\nu k_2^\mu-k_1\cdot k_2\; g^{\mu\nu}\Bigr)+\mathcal{F}_{2}\Bigl(k_3^\nu(k_1^\alpha k_2^\mu- k_1\cdot k_2\; g^{\alpha\mu})+k_2\cdot k_3(k_1^\nu g^{\alpha\mu}-k_1^\alpha g^{\mu\nu})\Bigr)\\
&+\frac{\mathcal{F}_{3}}{m_A^2}\;k_2^\alpha\Bigl(k_3^\mu(k_2\cdot k_3\;k_1^\nu-k_1\cdot k_2\; k^\nu)+k_1\cdot k(k_3^\nu k_2^\mu-k_2\cdot k_3\; g^{\mu\nu})\Bigr)+\Bigl(k_1^\mu\leftrightarrow k_2^\nu \Bigr),
\end{split}
\end{equation}
where  the form factors $\mathcal{F}_i$ depend on $s_1$, $s_2$, $s$, and $\mu_Z$, though we will refrain from writing out such a dependency explicitly. These form factors  will receive contributions from both  box and reducible diagrams, which means that the latter will  not generate  additional Lorentz structures.  We can thus write $\mathcal{F}_i=\mathcal{F}_i^{Box}+\mathcal{F}_i^{RD}$, where the notation is self-explanatory. The  expressions for the box diagram contributions  are too lengthy and they are presented in Appendix \ref{Amplitudes}   in terms of Passarino-Veltman scalar functions.

\subsubsection{Reducible diagram contribution}

\begin{figure}[ht!]
 \centering
 \includegraphics[width=8cm]{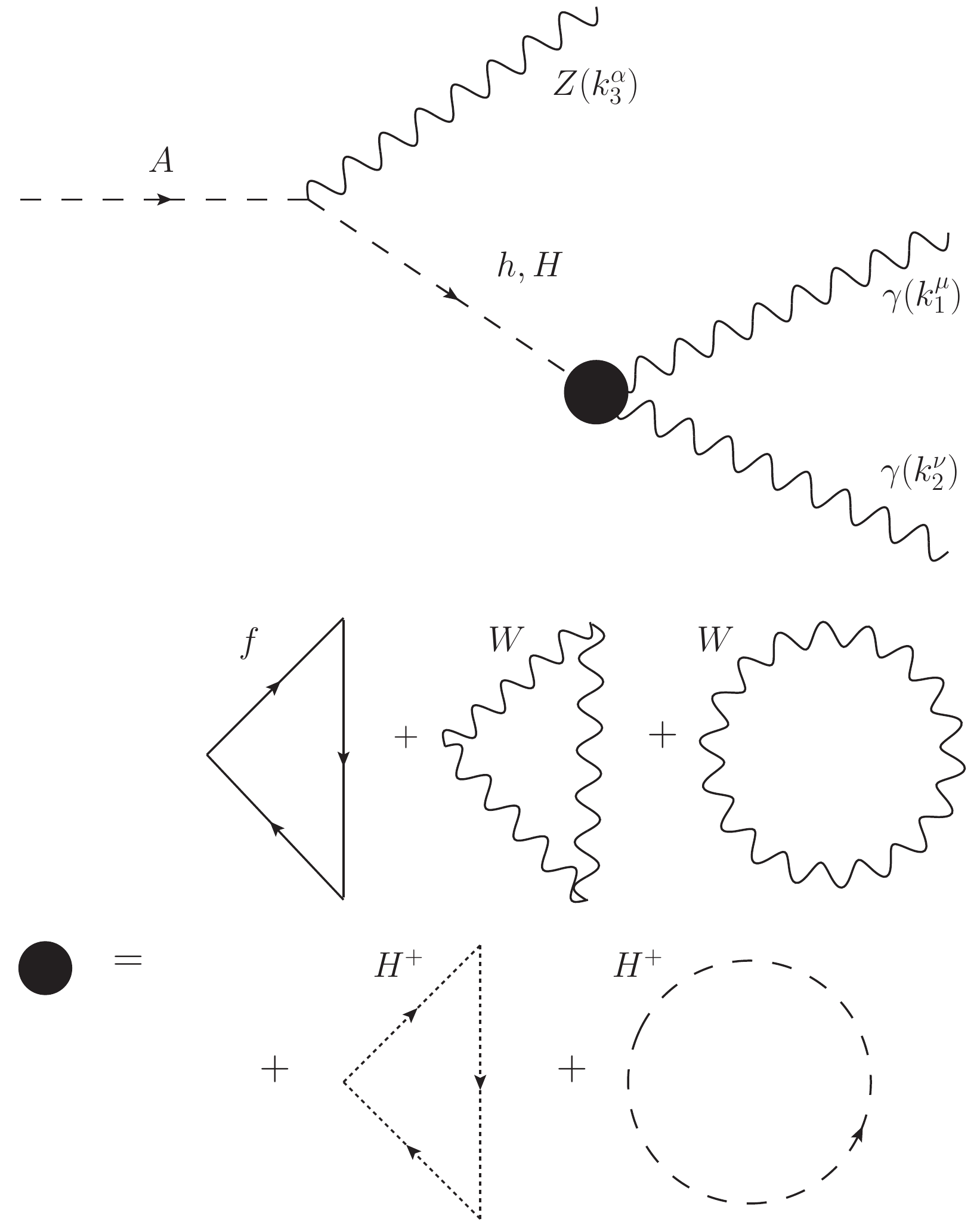}
 \caption{Reducible Feynman diagrams for the $A\to Z\gamma\gamma$ decay in the  THDM. For the triangle diagrams there are additional diagrams  that are obtained by  exchanging the photons.  Similar diagrams  also contribute to the $\phi \to Z\gamma\gamma$ ($\phi=h,H$) decay, except that the  intermediate particle is now the $CP$-odd scalar boson $A$ and there is only contribution from charged fermions in the triangle loop.}
 \label{FeynmanDiagramsAtoZggRed}
\end{figure}

There are also reducible diagrams in which the $A\to Z\gamma\gamma$ decay proceeds as  $A\to Z \phi^*\to Z\gamma\gamma$ ($\phi=h,H$),  as depicted in Fig. \ref{FeynmanDiagramsAtoZggRed}, with the two photons emerging from the intermediate scalar boson via loops carrying charged fermions,  the $W$ gauge boson, and the charged scalar boson $H^\pm$.

As was the case for the box diagram contribution, the reducible diagram contribution is  gauge invariant  and ultraviolet-finite by its own. It turns out that these diagrams contribute to the gauge-invariant amplitude of  Eq. \eqref{AtoZggAmplitude} only through the form factor ${\cal F}_1$, which includes the contributions of  charged fermions, the $W$ gauge boson, and the charged scalar boson $H^\pm$

\begin{equation}
\label{FRD}
\mathcal{F}_1^{RD}=\mathcal{F}_1^{f}+\mathcal{F}_1^{W}+\mathcal{F}_1^{H^\pm},
\end{equation}
with $\mathcal{F}_1^{\chi}$ ($\chi=f, {W}, {H^\pm}$) defined in Appendix \ref{FeynmanRules} in term of Passarino-Veltman scalar functions.

\subsection{ $\phi \to Z\gamma\gamma$ ($\phi=h,H$) decay}

\subsubsection{Box diagram contribution}
As for the $\phi \to Z    \gamma\gamma$ ($\phi=h,H$) decay, at the one-loop level it also receives the contributions of the fermion  box diagrams of Fig. \ref{FeynmanDiagramsHtoZggBox} with $A$ replaced by $\phi$. It is worth noting that although a $CP$-even scalar boson does couple to  charged $W^\mp$ gauge bosons and charged scalar bosons $H^\mp$, the corresponding box diagram contributions exactly cancel  out due to $CP$ invariance. Notice that the amplitude of this vertex must include the Levi-Civita tensor due to $CP$ invariance, but it cannot arise via box diagrams with charged particles other than charged fermions, whose coupling with the $Z$ gauge boson includes a $\gamma^5$ matrix. As  the invariant amplitude of a fermion loop includes the trace of a chain of Dirac matrices, the term involving the  $\gamma^5$ matrix would give rise to the required Levi-Civita tensor.

\begin{figure}[ht!]
 \centering
 \includegraphics[width=8cm]{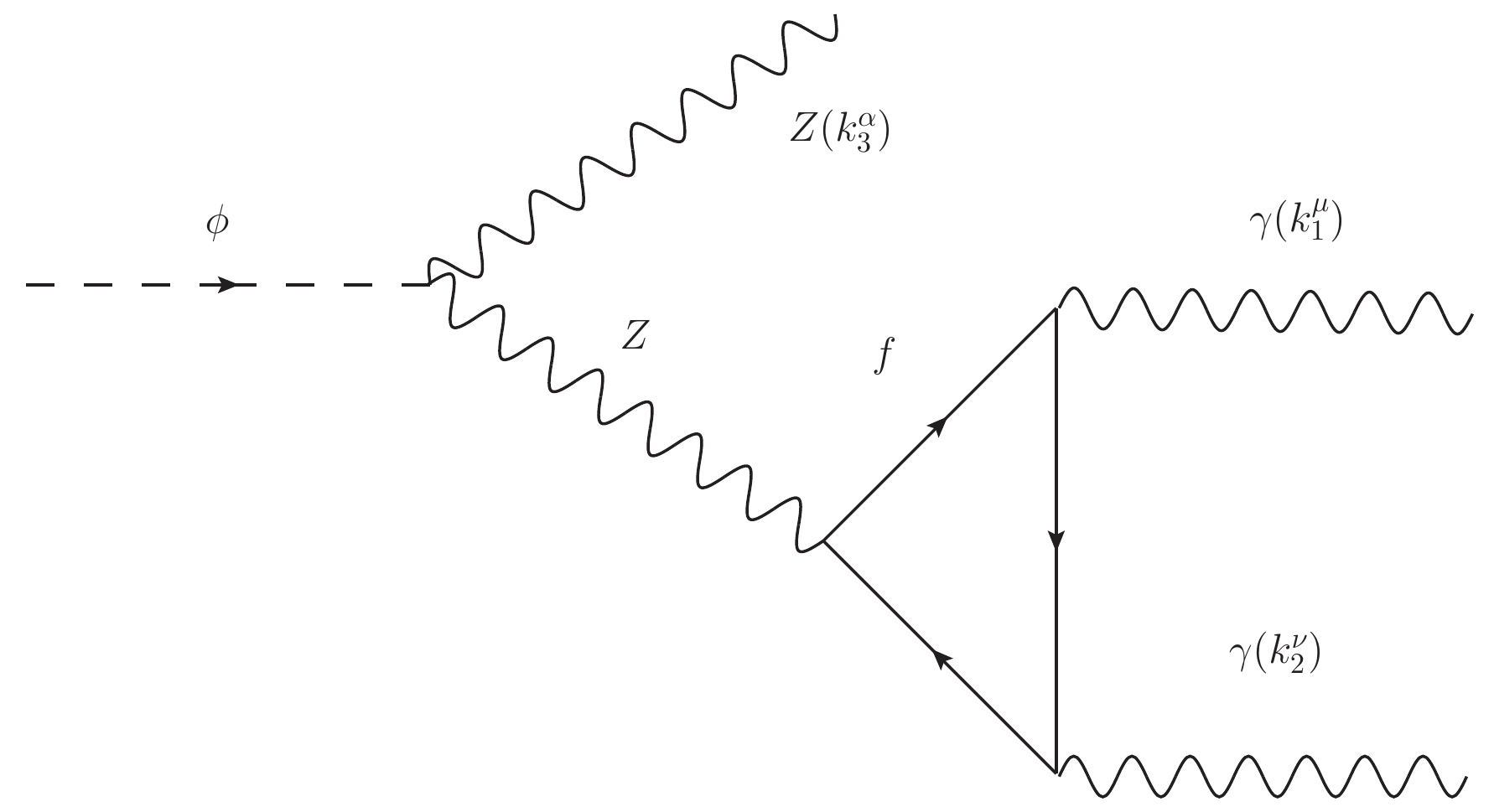}
 \caption{Feynman  diagram that also contributes to the $\phi\to Z\gamma\gamma$ ($\phi=h,H$) decay  in the  THDM, in addition to Feynman diagrams analogue to those of Figs. \ref{FeynmanDiagramsHtoZggBox} and \ref{FeynmanDiagramsAtoZggRed}. The diagram obtained by  exchanging the photons is not shown. }
 \label{FeynmanDiagramsHtoZggRed}
\end{figure}

The most general Lorentz structure for the  $\phi\to Z\gamma\gamma$ ($\phi=h,H$) decay can be written in the following gauge-invariant manifest form

\begin{equation}
\begin{split}
\label{phitoZggAmplitude}
\mathcal{M}^{\alpha\mu\nu}(\phi\to Z\gamma\gamma)&=\mathcal{G}_1 \Big(k_1\cdot k_2\;\epsilon^{\alpha\mu\nu k_3}+g^{\mu\nu}\epsilon^{\alpha k_3k_1k_2}-k_2^\mu\epsilon^{\alpha\nu k_3k_1}+k_1^\nu\epsilon^{\alpha\mu k_3 k_2}\Big)\\
&+\frac{\mathcal{G}_2}{m_\phi^2} \;\epsilon^{\alpha\mu k_3 k_1}\Big(k_3\cdot k_2\; k_1^\nu-k_1\cdot k_2 \;k_3^\nu\Big)+ \mathcal{G}_3 \Big(k_1\cdot k_2\;\epsilon^{\alpha\mu\nu k_1}+k_1^\nu\epsilon^{\alpha\mu k_1 k_2}\Big)\\
&+\mathcal{G}_4 \Big(k_3\cdot k_2\; \epsilon^{\alpha\mu\nu k_1}+k_3^\nu\epsilon^{\alpha\mu k_1 k_2}\Big)+\Big(k_1^\nu\leftrightarrow k_2^\mu\Big),
\end{split}
\end{equation}
where we use the shorthand notation $\epsilon^{\alpha k  p q}=\epsilon^{\alpha\beta\lambda\rho}k_\beta p_\lambda q_\rho$, etc.  Again the form factors $\mathcal{G}_i$ depend on $s$, $s_1$,  $s_2$, and $\mu_Z$. To arrive to the above equation, we used Schouten's identity. These form factors receive contributions from both box diagrams and reducible diagrams: $\mathcal{G}_i=\mathcal{G}_i^{Box}+\mathcal{G}_i^{RD}$. As far as the contributions from the box diagrams are concerned, they are reported in Appendix \ref{Amplitudes} in terms of Passarino-Veltman scalar functions.

\subsubsection{Reducible diagram contribution}
There are also contributions from reducible diagrams  that are analogue to those depicted in Fig. \ref{FeynmanDiagramsAtoZggRed}, but with the photons emerging from the intermediate $CP$-odd scalar boson $A$  via loops of charged fermions only.  There are also extra reducible diagrams arising from the process $\phi\to ZZ^*\to Z \gamma\gamma$, as shown in Fig. \ref{FeynmanDiagramsHtoZggRed}. This diagram involves the well-known triangle anomaly $Z^*\gamma\gamma$, which receives contributions from charged fermions only. This is due to $CP$ invariance  as the amplitude for this vertex must be proportional to the Levi-Civita tensor, which  can only arise via the trace of a chain of Dirac matrices including  $\gamma^5$, which in turn only is present in a fermion loop. Therefore loops of the charged $W$ gauge boson or the charged scalar boson do not contribute to this vertex. Also, due to the Landau-Yang theorem, the $Z^* \gamma\gamma$  vertex vanishes for real $Z$, so this diagram  does not contribute to the $\phi\to Z\gamma\gamma$ decay when the $\phi$ scalar boson is kinematically allowed to decay into a pair of real $Z$ gauge bosons. These reducible Feynman diagrams only  contribute to the invariant amplitude of the  $\phi\to Z\gamma\gamma$ ($\phi=h,H$) decay via the form factor $\mathcal{G}_3$:
\begin{equation}
\label{GRD}
\mathcal{G}_3^{RD}=\mathcal{G}_3^Z+\mathcal{G}_3^A,
\end{equation}
where $\mathcal{G}_3^Z$ and $\mathcal{G}_3^A$ are the form factors arising from  the diagrams with the vertices $Z^*\gamma\gamma$ and $A^*\gamma\gamma$, respectively. Explicit expressions in terms of Passarino-Veltman scalar functions are given in Appendix \ref{Amplitudes}. Note that we must include the contribution of all fermion families in order to cancel the  $Z^*\gamma\gamma$ anomaly.
\subsection{$A\to Z\gamma\gamma$ and $\phi\to Z\gamma\gamma$ ($\phi=h,H$) decay widths}

There are two scenarios for the    $\phi_i\to Z\gamma\gamma$ ($\phi=h,H,A$) decays, which depend on  the value of the mass of the incoming scalar boson $\phi_i$ as compared to the mass of  the exchanged scalar boson, which we denote  by $\phi_e$:  $\phi_e=h,H$ for $\phi_i=A$ or $\phi_e=A$ for $\phi_i=h,H$.  We will present the expression for the resulting decay width in both scenarios.

\subsubsection{$m_{\phi_i}<m_{\phi_e}+m_Z $}
In this scenario, the incoming scalar boson $\phi_i$ will not be heavy enough to produce an on-shell  $\phi_e$ in addition to the on-shell $Z$ gauge boson. Therefore  we will have a pure three-body decay induced by both  box  and reducible diagrams.  The corresponding decay width can be written as

\begin{equation}
\label{DecayWidth}
\Gamma(\phi_i\rightarrow Z\gamma\gamma) =\dfrac{m_{\phi_i}}{256\; \pi^3}\int_{x_{1i}}^{x_{1f}} \int_{x_{2i}}^{x_{2f}} |\overline{\mathcal{M}}(\phi_i\rightarrow Z\gamma\gamma) |^2 dx_2\;dx_1,
\end{equation}
where we introduced the following scaled variables
\begin{align}\label{scaledvariables}
x_1& = \frac{2p\cdot k_3}{m_{\phi_i}^2}=1+\mu_Z-\hat s,\\
x_2 & = \frac{2p\cdot k_1}{m_{\phi_i}^2}=1-\hat s_2,\\
x_3 &= \frac{2p\cdot k_2}{m_{\phi_i}^2}=1-\hat s_1,
\end{align}
with $\hat s=s/m_{\phi_i}^2$ and $\hat s_i=s_i/m_{\phi_i}^2$. In the center-of-mass frame of the decaying $\phi_i$ we have $x_1=2E_Z/m_{\phi_i}$, $x_2=2E_\gamma/m_{\phi_i}$, and $x_3=2E_{\gamma'}/m_{\phi_i}$, where $E_\gamma$ ($E_{\gamma'}$) stands for the energy of the photon with four-momentum $k_1$ ($k_2$). From energy conservation, these variables  obey $x_1+x_2+x_3=2$.

The kinematic limits in  Eq. \eqref{DecayWidth} are as follows
\begin{equation}
\begin{split}
x_{1i} &=2\sqrt{\mu_Z},\\
x_{1f}&=1+\mu_Z,\\
x_{2i,2f}&=\dfrac{1}{2}\left (2-x_1\mp \sqrt{x_1^{ 2}-4\mu_Z}\right ).
\end{split}
\end{equation}
The squared average amplitudes for both  decays $A\to Z\gamma\gamma$ and $\phi\to Z\gamma\gamma$ ($\phi=h,H$) are presented in Appendix \ref{SquaredAmplitudes}.

\subsubsection{$m_{\phi_i}> m_{\phi_e}+m_Z $}

In this  scenario the incoming scalar boson $\phi_i$  is heavy enough to produce an on-shell scalar boson $\phi_e$. Therefore the $\phi_i\to Z\gamma\gamma$ decay proceeds as the pure two-body decay $\phi_i\to Z\phi_e$, followed by the decay $\phi_e\to \gamma\gamma$. Note that in the case of the decay of a $CP$-even Higgs boson, although the decay into  a pair of real $Z$ gauge bosons will now be kinematically allowed, the $Z\to\gamma\gamma$ decay is forbidden by the Landau-Yang theorem, which means that the contribution of the intermediary $Z$ gauge boson will thus vanish.  In this scenario, using the Breit-Wigner propagator for the exchanged scalar boson, Eq. \eqref{DecayWidth} can be integrated and  the $\phi_i\to Z\gamma\gamma$ decay width can be written as

\begin{equation}
\Gamma(\phi_i\to Z\gamma\gamma)=\Gamma(\phi_i \to Z\phi_e)\mbox{BR}(\phi_e\to\gamma\gamma),
\end{equation}
with the  decay width $\Gamma(\phi_e\to \gamma\gamma)$  given by
%
%\begin{equation}
% \Gamma(\phi\to \gamma\gamma)=\dfrac{\alpha^2}{8m_\phi\pi^3}\left |\mathcal{F}_1+\mathcal{F}_2\right |^2,
% \end{equation}
%with
%
%\begin{eqnarray}
%\mathcal{F}_1&=&\sum_f m_fN_c^fg_{\phi ff}((4m_f^2-m_{\phi}^2)C_1(m_\phi^2)+2),\\
%\mathcal{F}_2&=&\dfrac{g_{\phi WW}}{4m_W^2}\Big(6m_W^2(m_\phi^2-2m_W^2)C_1^\prime(m_\phi^2)-m_\phi^2 -6m_W^2\Big).
%\end{eqnarray}
\begin{equation}
\label{phitoggdecaywidth}
\Gamma(\phi_e\to \gamma\gamma)=\frac{\alpha^2g^2m_\phi^3}{1024\pi^3m_W^2}\left|\mathcal{F}^{\phi_e\gamma\gamma}\right|^2.
\end{equation}
For a $CP$-even scalar boson $\phi_e=h,H$, $F^{\phi_e \gamma\gamma}$ receive contributions from  charged fermions, the charged $W$ gauge boson, and the charged scalar boson $H^\pm$:
\begin{equation}
\label{phitoggFiTot}
\mathcal{F}^{\phi_e\gamma\gamma} =\mathcal{F}^{\phi_e\gamma\gamma}_f(\tau_f)+\mathcal{F}^{\phi_e\gamma\gamma}_W(\tau_W)+\mathcal{F}^{\phi_e\gamma\gamma}_{H^\pm}(\tau_{H^\pm})\,\quad {\rm for}\quad \phi_e=h,H,
\end{equation}
 with $\tau_\chi=4 m_\chi^2/m_{\phi_e}^2$. The $\mathcal{F}^{\phi_e \gamma\gamma}_\chi(x)$ functions can be obtained from the results for the reducible diagrams presented in Appendix \ref{Amplitudes} by setting $s=m_{\phi_e}^2$. They are given by
\begin{equation}
\label{phitoggFi}
\mathcal{F}^{\phi_e \gamma\gamma}_\chi (\tau_\chi)=\left\{
\begin{array}{cl}
\sum_f g_{\phi_e \bar{f}f} N_c^f Q_f^2\left[-2\tau_f(1+(1-\tau_f)f(\tau_f))\right]&\quad \chi=f,\\ \\
  g_{\phi_e WW}\left[2+3\tau_W+3\tau_W(2-\tau_W)f(\tau_W)\right]&\quad \chi=W,\\ \\
 \dfrac{m_W^2 (1-2s_W^2) g_{\phi_e H^-H^+}}{c_W^2 m_{H^\pm}^2}\left[\tau_{H^\pm}(1-\tau_{H^\pm}f(\tau_{H^\pm}))\right]&\quad \chi=H^\pm,
 \end{array}\right.
\end{equation}
for $\phi_e=h,H$ $f(x)$ is given by
\begin{equation}
\label{f(x)}
f(x)=\left\{
\begin{array}{cr}
\left[\arcsin\left(\frac{1}{\sqrt{x}}\right)\right]^2&x\ge1,\\
-\frac{1}{4}\left[\log\left(\frac{1+\sqrt{1-x}}{1-\sqrt{1-x}}\right)-i\pi\right]^2&x<1.
\end{array}
\right.
\end{equation}

On the other hand, when the intermediary scalar boson is the $CP$-odd one $A$, we only have the contributions of charged fermions
\begin{equation}
\label{AtoggFiTot}
\mathcal{F}^{A\gamma\gamma}=\mathcal{F}^{A\gamma\gamma}_f(\tau_f)=
\sum_f g_{A\bar{f}f} Q_f^2 N_c^f\left[-2\tau_f f(\tau_f)\right].
\end{equation}
As for the  $\phi_i \to \phi_e Z$ decay width, it is given as follows

\begin{equation}
\label{phitoAZ}
\Gamma(\phi_i \to Z\phi_e )=\frac{g_{\phi Z\phi_e}^2\,\alpha\, m_{\phi_i}^3}{256 s_W^2
}\Bigl((4-(\sqrt{\tau_{\phi_e}}-\sqrt{\tau_Z})^2)(4-(\sqrt{\tau_{\phi_e}}+\sqrt{\tau_Z})^2)\Bigr)^{\frac{3}{2}}.
\end{equation}
Note that $\tau_{\phi_e}=4 m_{\phi_e}^2/m_{\phi_i}^2$, thus $\tau_{\phi_e}=4 m_{\phi}^2/m_A^2$ for $A\to Z\gamma\gamma$ and  $\tau_{\phi_e}=4 m_{A}^2/m_\phi^2$ for $\phi \to Z\gamma\gamma$ ($\phi=h,H$). A similar expression with the corresponding replacements is obeyed by the $\phi_e\to W^\pm H^\mp$ decays if kinematically allowed.

All the necessary coupling constants $g_{\phi \bar{f}f}$, $g_{\phi WW}$, $g_{\phi ZZ}$, $g_{\phi AZ}$, $g_{\phi W^\pm H^\mp}$, $g_{\phi H^-H^+}$ $(\phi=h,H)$, along with $g_{A\bar{f}f}$ and   $g_{A W^\pm H^\mp}$ are shown in  Table \ref{CouplingConstants} of Appendix \ref{FeynmanRules}. Other coupling constants involved in decays  such as  $H\to hh$ and $\phi\to AA$ can be found in Ref. \cite{Branco:2011iw,Gunion:1989we} for instance. To obtain the    branching ratio   $BR(\phi_e \to \gamma\gamma)$, we need  the main decay widths  of both  $CP$-even and $CP$-odd scalar bosons, which have already been studied in the literature considerably \cite{Gunion:1989we}. For completeness we present in Appendix \ref{DecayWidthFormulas} all the necessary formulas, which can also be helpful to obtain the branching ratio for the $\phi_i \to Z\gamma\gamma$ decay in type-II THDM and make a comparison with that of other decay channels.

\section{Numerical Analysis and Results}
We now turn to the numerical analysis. To begin with, we will analyze the current constraints on the parameter space of type-II THDM.

\subsection{Allowed parameter space of type-II THDM}

After the Higgs boson discovery, several studies have been devoted to explore the implications on the parameter space of THDMs \cite{Grinstein:2013npa,Dumont:2014wha,Dorsch:2016tab,Han:2017pfo,Eberhardt:2013uba}. From the recent analyses of the ATLAS and CMS collaborations \cite{Khachatryan:2016vau}, it is inferred that   the properties of the $125$ GeV scalar boson found at the LHC  are  highly consistent with the SM predictions, thereby imposing strong constraints on the scalar sector of SM extensions. If one of such theories predicts several $CP$-even physical Higgs bosons, one of them  must correspond to the SM one and  reproduce its  couplings to fermions and gauge bosons. In type-II THDM, the scalar boson $h$ is usually assumed to be the lightest one and so is identified with the SM Higgs boson, which constrains the parameter space of the model to a region very close  to the alignment limit  $\sin(\beta-\alpha)= 1$, where the  heavy Higgs $H$ does not couple to the gauge bosons and the coupling $hZA$ is absent at  tree-level \cite{Aad:2015pla,Eberhardt:2013uba,Pich:2009sp}. The couplings of the  $h$ Higgs boson  to the fermions  involve the mixing angles $\alpha$ and $\beta$. Therefore, the LHC data   can impose  strong constraints on both parameters. Other constraints can be obtained from theoretical requirements such as  vacuum stability and unitarity of the scalar potential as well as perturbativity of the Higgs couplings. Also, the oblique parameters $S, T$ and $U$ can  impose  strong constraints on the masses of the new Higgs bosons $A$ and $H$,  requiring that  at least one of them is very heavy: a $CP$-odd scalar with $m_A\sim 200$ GeV requires a  heavy $CP$-even scalar with $m_H\ge 600$ GeV and viceversa. As for the charged Higgs boson mass, it can be constrained through experimental measurements on low energy FCNC processes.

All of the above constraints can be complemented with the direct searches of additional Higgs bosons at LEP and the LHC. Below we present  the constraints most relevant for our numerical analysis.

\begin{itemize}
\item Mixing angles $\beta$ and $\beta-\alpha$:  since the $h$ Higgs boson is identified with the SM Higgs boson, the LHC data restrict $\beta-\alpha$   to  lie very close to $\pi/2$, namely, $|\sin(\beta-\alpha)|> 0.999$, with a small interval around $t_\beta=1$ where such a constraint is less stringent. Furthermore, in type-II THDM, for $\beta-\alpha\simeq \pi/2$, the scalar couplings to the top quark (bottom quark) behaves as $1/t_\beta$ ($t_\beta$), thus FCNCs process are very sensitive to small and large values of $t_\beta$, which will impose stringent constraints on this parameter. We can thus consider values of $t_\beta$ in the range 1-30.

\item Mass of the charged Higgs boson $m_{H^\pm}$:  while the direct search at LEP imposed the constraint $m_{H^\pm} > 80$ GeV \cite{Abbiendi:2013hk}, the measurement of the $\bar B \to X_s\gamma$ branching ratio imposes the very stringent bound  $m_{H^\pm} > 570$ GeV, for $t_\beta\sim 1.5$  \cite{Misiak:2017bgg}.

\item Mass of the $CP$-odd scalar $m_A$: the authors of Ref.  \cite{Han:2017pfo} examine the scenarios where either $m_A$ or $m_H$ is set to a large value about 600-700 GeV while the other one is bounded via theory constraints and experimental data, with the remaining free parameters set to the values mentioned above. We will  follow closely this analysis  as it is of interest for the present work. We first examine the case of a light $CP$-odd scalar and a heavy $CP$-even scalar with mass  $m_H = 600$ GeV. In this scenario the searches for the decays  $A\to \bar{\tau}\tau$, $A\to\gamma\gamma$, and $A\to hZ$  exclude the region  $m_A < 350$ GeV, whereas the LHC data on the Higgs boson  require $m_A$ to be larger than $220 $ GeV. On the other hand, the search for the channel $b\bar{b} \to A \to \bar{\tau}\tau$ allows for $m_A$ values  in the range  350-700 GeV and impose the upper limit $t_\beta< 2$ for $m_A\le 500$ GeV, whereas $t_\beta< 15$ for $m_A\le $ 500 GeV.

\item Mass of the heavy $CP$-even scalar $m_H$:  we now examine the scenario with a light $CP$-even scalar and a heavy $CP$-odd scalar  with $m_A=700$ GeV. In this case the whole  constraints require $m_H>300$ GeV, whereas  the $b\bar{b}\to H/A \to \tau\bar{\tau}$ channel imposes an  upper bound on $t_\beta$ as a function of $m_H$. For instance, for $m_H = 200$ (600) GeV $t_\beta < 6$ (15). On the other hand,  the searches for the $H$ decays into $\bar{\tau}\tau$, $WW$, $ZZ$, $\gamma\gamma$, and $hh$ require $\tan\beta > 2.5$ for $m_H < 380$ GeV. For a lighter $m_A = 600$  GeV, the search for the $A\to HZ$ channel can exclude the region $m_H < 270$ GeV.
\end{itemize}

%In summary we can choose the following values for the model parameters involved in our calculation:

We now turn to study the behavior of the $A\to Z\gamma\gamma$ and $\phi \to Z\gamma\gamma$ ($\phi=h,H$)  branching ratios as functions  of the parameters $t_\beta$, $\beta-\alpha$, $m_{H^\pm}$, $m_A$ and $m_H$. We stick to the still allowed values for these parameters, whereas for the SM parameters we take the values given in Ref. \cite{Olive:2016xmw}. For our analysis we used the LoopTools package \cite{vanOldenborgh:1989wn,Hahn:1998yk} for the numerical evaluation of the Passarino-Veltman scalar functions appearing in the decay amplitudes. The dominant decay widths of the $CP$-odd and $CP$-even scalar bosons were evaluated by our own Mathematica code that implements the formulas of Appendix \ref{DecayWidthFormulas}, including the QCD corrections for the decays into light quarks.

\subsection{$A\to Z\gamma\gamma$   branching ratio}

We work in a region close to the alignment limit and use  $\sin(\beta-\alpha)=0.999$. In this scenario, the  strength of the $hZA$ vertex is negligible and so the contributions to the $A\to Z\gamma\gamma$  decay only arise from  box diagrams and reducible diagrams with $H$ exchange, which receive their main contributions  from the top and bottom quarks. The contribution of the loops with $W$  gauge bosons turns out to be negligibly small as it is proportional to $\cos^2(\beta-\alpha)$, whereas the charged scalar boson  also gives a very small contribution for  $m_{H^\pm}$ of the order of a few hundred GeVs.

We can distinguish two scenarios of interest: $m_A< m_H+m_Z$ and $m_A> m_H+m_Z$. Below we examine the behavior of the $A\to Z\gamma\gamma$ branching ratio in such scenarios.

\subsubsection{Scenario with $m_A< m_H+m_Z$ }

\begin{figure}[ht!]
 \includegraphics[width=0.9\textwidth]{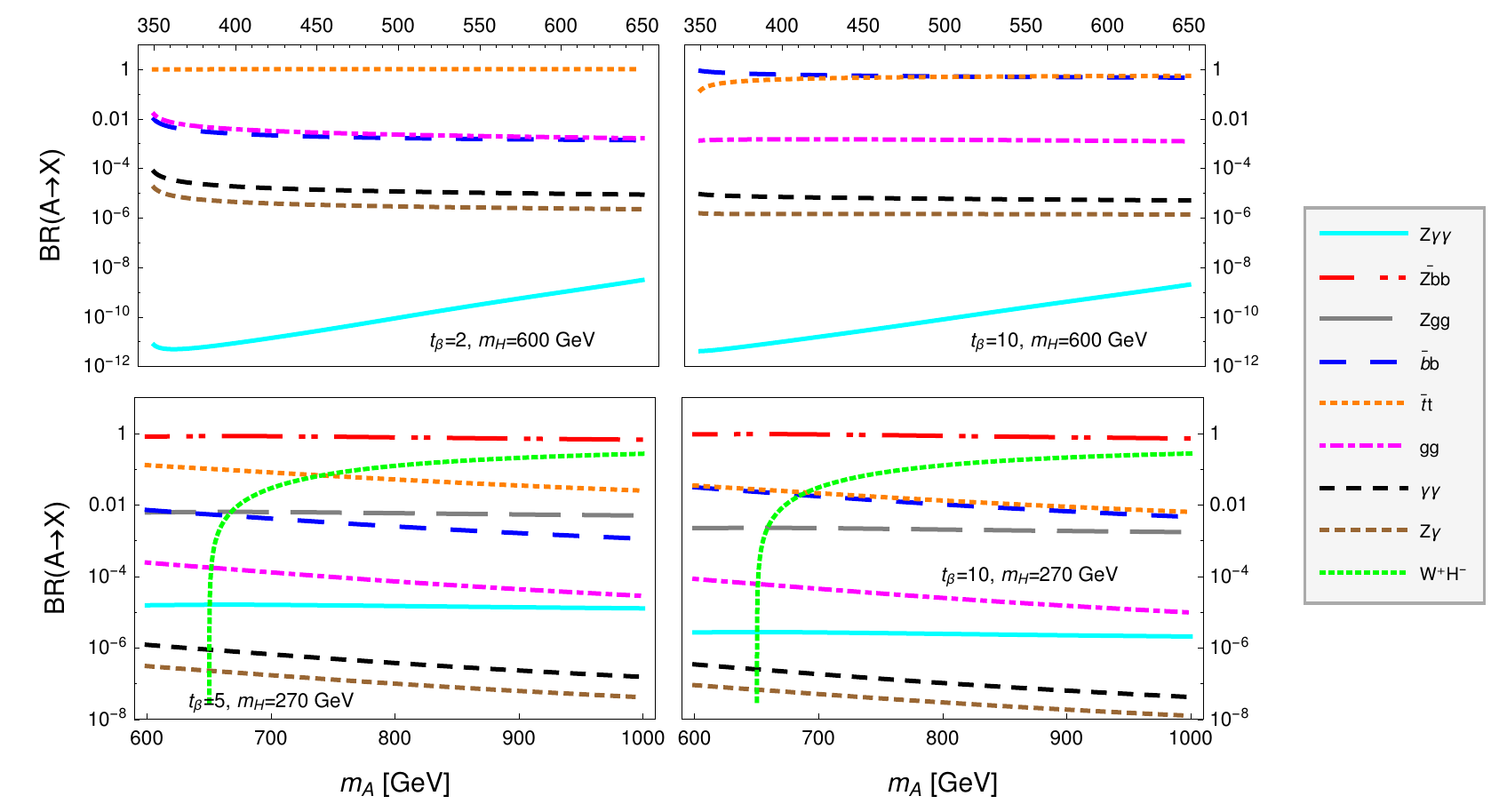}
\caption{Branching ratio for the $A\to Z\gamma \gamma$ decay in type-II THDM as a function of $m_A$ for $m_{H^\pm}=570$ GeV, $\sin(\beta-\alpha)=0.999$, and two values of $t_\beta$ allowed by theory and experimental constraints. In the upper (lower) plots we use $m_H=600$  (270) GeV. The branching ratios for the main $A$ decay channels are also shown.\label{BRAtoZgg}}
\end{figure}

We consider the scenario with $m_H=600$ GeV and analyze the behavior of $BR(A\to Z\gamma\gamma)$ as a function of $m_A$ in the range $350-650$ GeV. For the mixing angle $\beta$ we consider two values: $t_\beta=2$ and $t_\beta=10$, which are allowed  for $m_A<500$ GeV and 500 GeV $\le m_A\le 700$ GeV, respectively. In the upper plots of Fig. \ref{BRAtoZgg} we show the behavior of the $A\to Z\gamma\gamma$ branching ratio as function of  $m_A$ for the two chosen values of $t_\beta$. We also show the main decay modes of the $CP$-odd scalar boson: $A\to b\bar{b}$, $A\to W^-H^+ $, $t\bar{t}$, $gg$, $\gamma\gamma$, and $Z\gamma$. The decay $A\to Zh$ has a negligible branching ratio in the region close to the alignment limit and  is not shown in the plots. We note that the main contribution to $BR(A\to Z\gamma\gamma)$ arises from the reducible diagrams with  top quarks, whereas the contribution of the loops with  charged scalar bosons  is negligible. Since in this scenario the intermediary scalar boson $H$ is far from  the resonance,  the reducible diagram contribution is very small, though it is larger than the box diagram contribution by almost two orders of magnitude. Therefore, the $Z\to A\gamma\gamma$ branching ratio is thus very small. For instance, for $t_\beta=2$, $BR(A\to Z\gamma\gamma)$ is of the order of $10^{-11}$ for $m_A=300$ GeV with a small increase as $m_A$ increases. When $t_\beta$ increases up to $10$, $BR(A\to Z\gamma\gamma)$ decreases about one order of magnitude as the top quark contribution is suppressed by a factor of $1/t_\beta$. In this region of the parameter space of the type-II THDM, the $A\to Z\gamma \gamma$  branching ratio is considerably smaller than those of the one-loop induced decays $A\to \gamma\gamma$ and $A\to Z\gamma$.

\subsubsection{Scenario with $m_A> m_H+m_Z$ }
We now turn to analyze the scenario where the  $CP$-even scalar is relatively light,  with a mass $m_H=270$ GeV along with a heavier $CP$-odd scalar with a mass in the range 600-1000 GeV.  We use $t_\beta=5$ and $t_\beta=10$, which are allowed for  $m_A=600$ GeV and $m_A=700$ GeV, respectively. In this scenario the intermediate scalar boson $H$ is on resonance and  the $CP$-odd scalar can decay as $A\to ZH$ with a large branching ratio.  The decay $A\to Z\gamma\gamma$ would then proceed in two stages: after the $CP$-odd scalar boson decays as $A\to ZH$,  the on-shell $CP$-even scalar boson decays into a photon pair $H\to \gamma\gamma$, namely, $A\to H Z\to Z\gamma\gamma$. The enhancement of  $BR(A\to Z\gamma\gamma)$  becomes evident in the lower plots of Fig.  \ref{BRAtoZgg}, where we show its behavior as a function of $m_A$, along with that of the branching ratios of other decay modes of the $CP$-odd scalar boson. We observe that $BR(A\to Z\gamma\gamma)$ increases up to four orders of magnitude with respect to the result obtained in the scenario with $m_A< m_H+m_Z$ and can reach values of the order of $10^{-6}-10^{-5}$ when  $m_A$ is in the  600-800 GeV range. In this mass regime, the main decay is $A\to HZ$, which explains why the  $A\to Z\gamma\gamma$ decay has such an enhanced branching ratio. For illustrative purpose we also show the branching ratios for the decays $A\to Z\bar{b}b$ and  $A\to  Zgg$, which arise from the decay $A\to HZ$ followed by the decays $H\to\bar{b}b$ and $H\to gg$. We note that the dominant decay channel is   $A\to Z\bar{b}b$.

\begin{figure}[ht!]
 \includegraphics[width=0.45\textwidth]{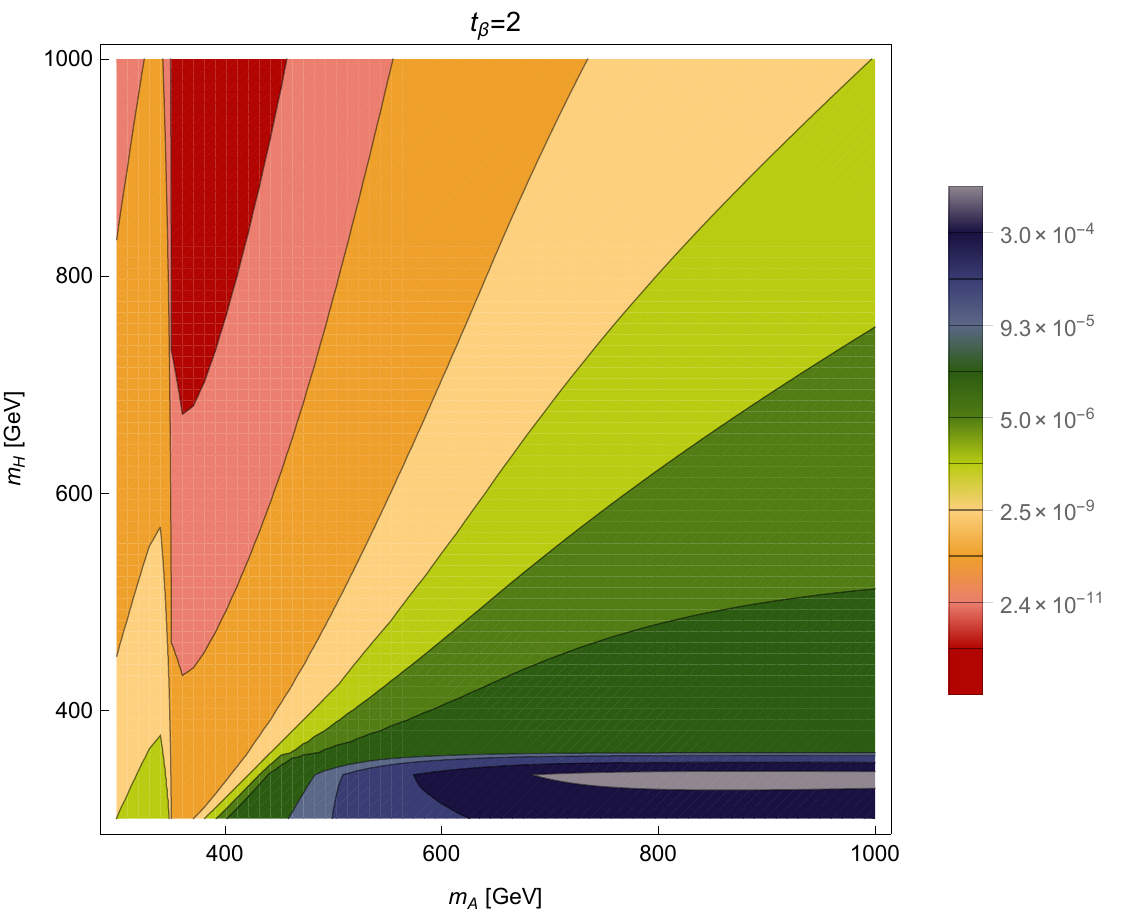}
\includegraphics[width=0.45\textwidth]{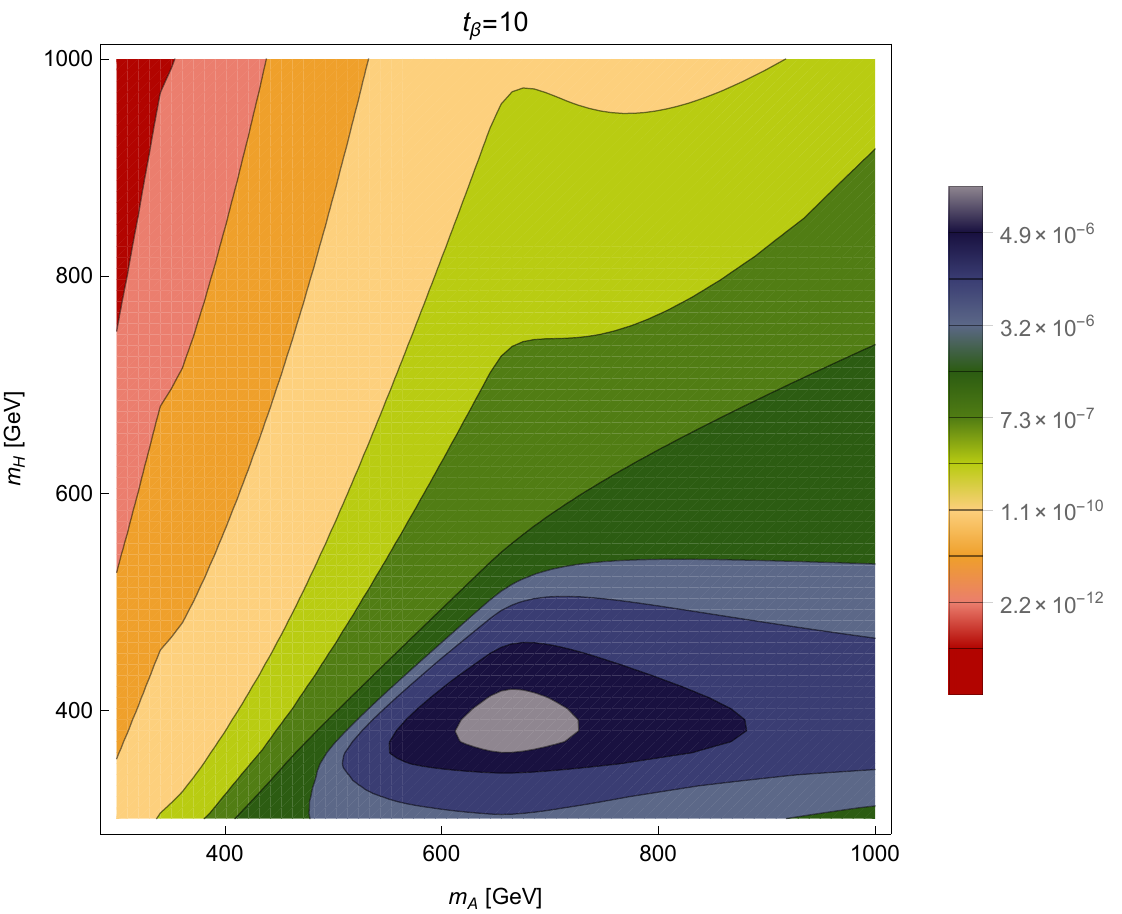}
\caption{Contour plot of  $BR(A\to Z\gamma\gamma)$ in the $m_A$ vs $m_H$ plane for $\sin(\beta-\alpha)=0.999$, $m_{H^\pm}>m_A$ and  two values of $t_\beta$.\label{BRAtoZggContour}}
\end{figure}

The above-described behavior of $BR(A\to Z\gamma\gamma)$ is best illustrated in the contour plot on the $m_A$ vs $m_H$ plane shown in Fig. \ref{BRAtoZggContour} for two values of $t_\beta$. We observe that $BR(A\to Z\gamma\gamma)$ can reach its largest values, of the order of $10^{-5}$, in the region where $m_A>m_A+m_Z$, whereas it is negligible when $m_A<m_A+m_Z$. Since  the $A\to Z\gamma\gamma$ decay receives its main contribution from the loops with top quark, it decreases as $t_\beta$ increases.

\subsection{$H\to Z\gamma\gamma$  branching ratio}
We now   analyze the behavior of  the  $H\to Z\gamma\gamma$ branching ratio as a function of $m_H$ in scenarios analogue to those discussed for the $CP$-odd scalar boson. For $\sin(\beta-\alpha)=0.999$, apart from the box diagram contribution, the only contribution from reducible diagrams is that with an intermediary $CP$-odd scalar boson $A$, which receives contributions mainly from the top and bottom quarks. The diagram mediated by the $Z$ gauge boson gives a negligible contribution since the $HZZ$ vertex is proportional to $\cos(\beta-\alpha)$.

\subsubsection{Scenario with $m_H< m_A+m_Z$ }
\begin{figure}[ht!]
 \includegraphics[width=0.9\textwidth]{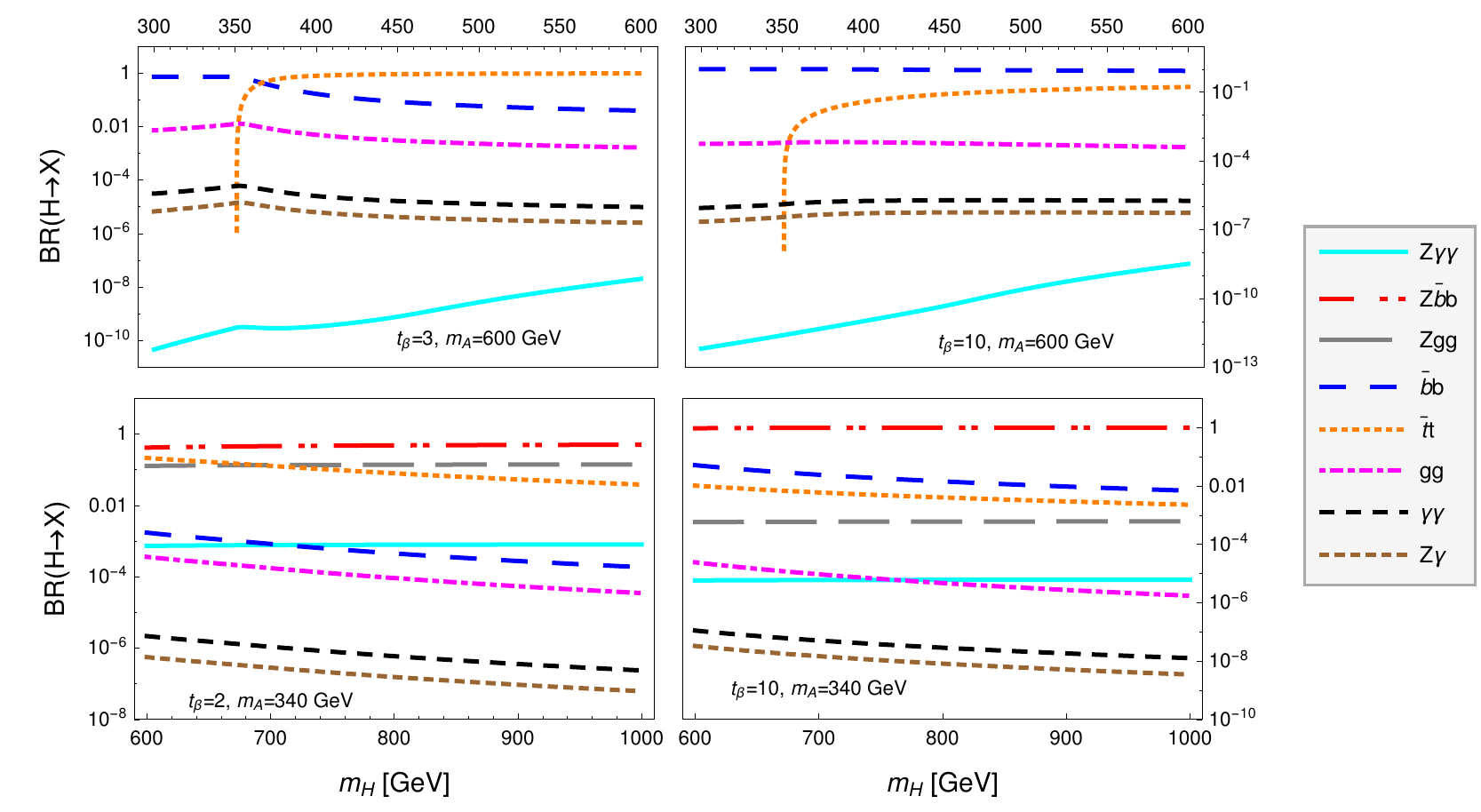}
\caption{Branching ratio for the $H\to Z\gamma \gamma$ decay in type-II THDM as a function of $m_H$ for $m_{H^\pm}=570$ GeV, $\sin(\beta-\alpha)=0.999$, and two values of $t_\beta$ allowed by theory and experimental constraints.  In the upper (lower) plots we use $m_A=600$  (350) GeV.   The branching ratios for the main $H$ decay channels are also shown.\label{BRHtoZgg}}
\end{figure}

We consider a heavy $CP$-odd scalar with a mass $m_A=600$ GeV and take $m_H$ in the range 300-600 GeV. For $t_\beta$ we  use the values 3 and 10. In the upper plots of Fig. \ref{BRHtoZgg} we show the  branching ratios for the main decay channels of the $H$ scalar boson. We note that the $H\to Z \gamma\gamma$ decay has a very suppressed branching ratio up to five orders of magnitude smaller than the branching ratios of the one-loop induced decays $H\to\gamma\gamma$ and $H\to Z\gamma$. It increases for smaller $t_\beta$ but it seems still beyond the reach of detection.

\subsubsection{Scenario with $m_H> m_A+m_Z$ }
In this scenario we consider $m_A=350$ GeV and   take $m_H$ in the range 600-1000 GeV. We  also use $t_\beta=2$ and $t_\beta=10$. For the mass of the charged scalar boson we use $m_{H^\pm}=575$ GeV as we do not need to assume that $m_{H^\pm}> m_H$ since the $H\to W^-H^+$ decay channel has a negligible branching ratio proportional to $\cos(\beta-\alpha)^2$. In the lower plots of Fig. \ref{BRHtoZgg} we show the $H\to Z\gamma\gamma$ branching ratio along with those of the main decay channels. We note that there is a considerable enhancement of $BR(H\to Z\gamma\gamma)$, up to 5 orders of magnitude,  now that the $H\to ZA $ decay is allowed, thus $BR(H\to Z\gamma\gamma)$ can be as large as $10^{-3}$ for $t_\beta=2$. Again we include the decays $H\to ZA\to Z\bar{b}b$ and  $H\to ZA\to Zgg$, with the dominant decay being the  $H\to ZA\to Z\bar{b}b$.

\begin{figure}[ht!]
 \includegraphics[width=0.45\textwidth]{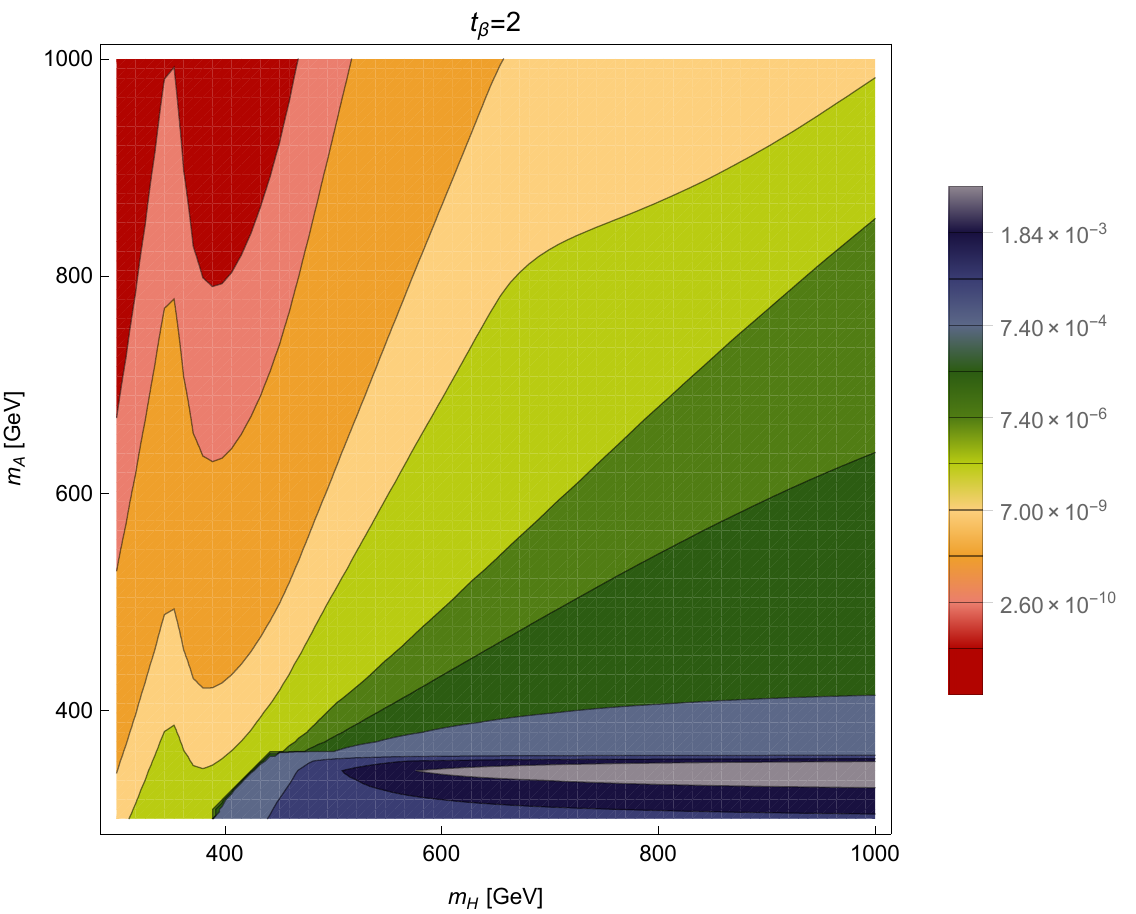}
\includegraphics[width=0.45\textwidth]{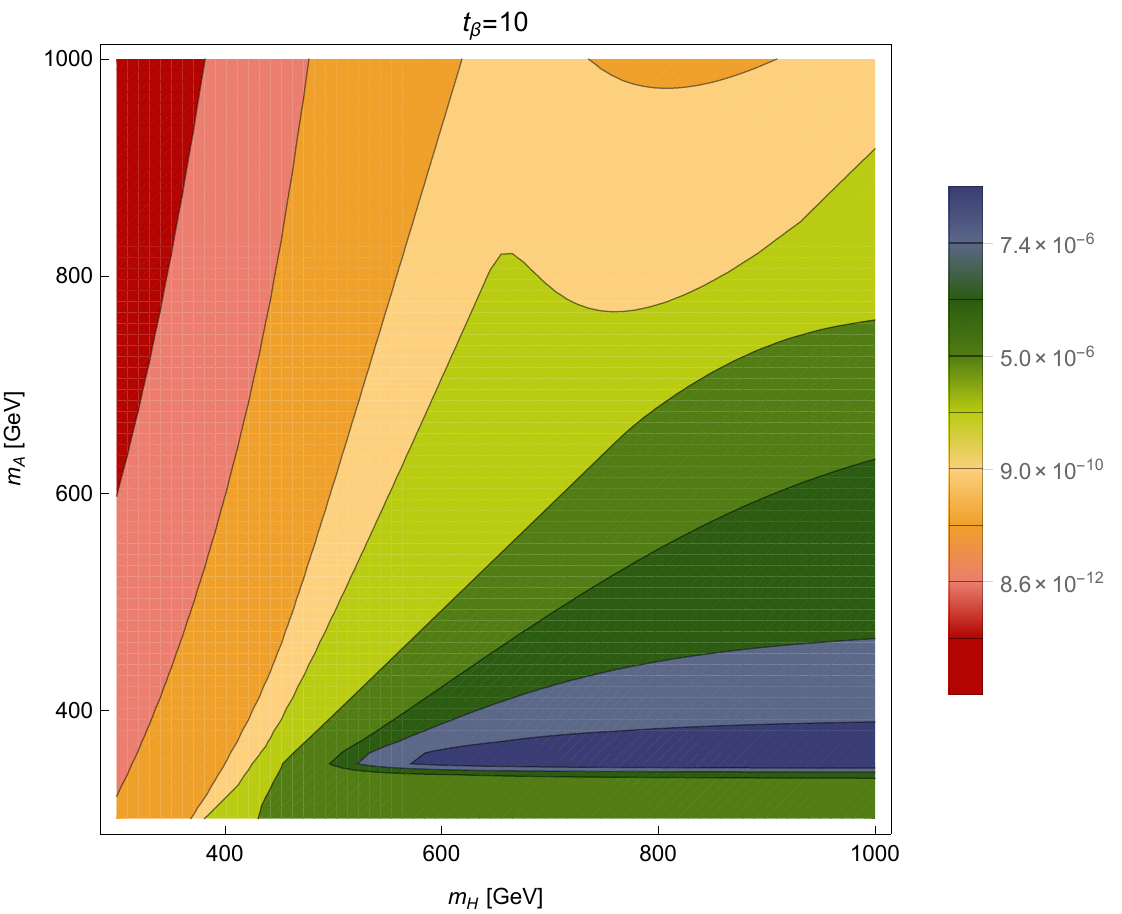}
\caption{Contour plot of  $BR(H\to Z\gamma\gamma)$ in the $m_H$ vs $m_A$ plane for $\sin(\beta-\alpha)=0.999$, $m_{H^\pm}=570$ GeV and  two values of $t_\beta$.\label{BRHtoZggContour}}
\end{figure}

 In Fig. \ref{BRHtoZggContour}, we also show the contour plot of $BR(H\to Z\gamma\gamma)$ in the $m_H$ vs $m_A$ plane for two values of $t_\beta$. Again it is evident that $BR(H\to Z\gamma\gamma)$ can reach its largest values when $m_H>m_A+m_Z$ and it is negligible when $m_H<m_A+m_Z$. It also decreases when $t_\beta$ increases as it receives the main contribution from loops with the top quark.

Finally, we would like to comment shortly on the potential detection of the  $A\to Z\gamma\gamma$ and $\phi \to Z\gamma\gamma$ ($\phi=h,H$) decays in the scenario we are considering in type-II THDM. Although there is a considerable enhancement of the corresponding branching ratios, it still seems not enough to put these decays at the reach of experimental detection at the LHC in the near future. In Fig.  8 we show the leading order production cross section for the $CP$-even and $CP$-odd  scalar bosons via gluon fusion at the LHC at $\sqrt{s}=$ 14 TeV as a function of the scalar boson mass. It turns out that with an integrated luminosity of 300 fb$^{-1}$, to be achieved in LHC run 3, we would have about $1.64 \times 10^5$ ($3.2 \times 10^5 $) $CP$-even ($CP$-odd) scalar bosons with a mass $m_\phi=500$ GeV produced per year, but these numbers drop by one order of magnitude when $m_\phi=700$ GeV. For $BR(H\to Z\gamma\gamma)\simeq O(10^{-3})$, we would only have about 164  $H\to Z\gamma\gamma$ events prior to imposing the kinematic cuts, which would render this decay hard to detect. The situation might be more promising  at a future high-luminosity 100 TeV $pp$ collider, where we could have thousands of $H\to Z\gamma\gamma$ events prior to imposing the kinematic cuts. As  discussed below, this event number would increase in type-I THDM by one order of magnitude as the respective branching ratios would have such an enhancement in that model.

\begin{figure}
  \centering
  \includegraphics[width=10cm]{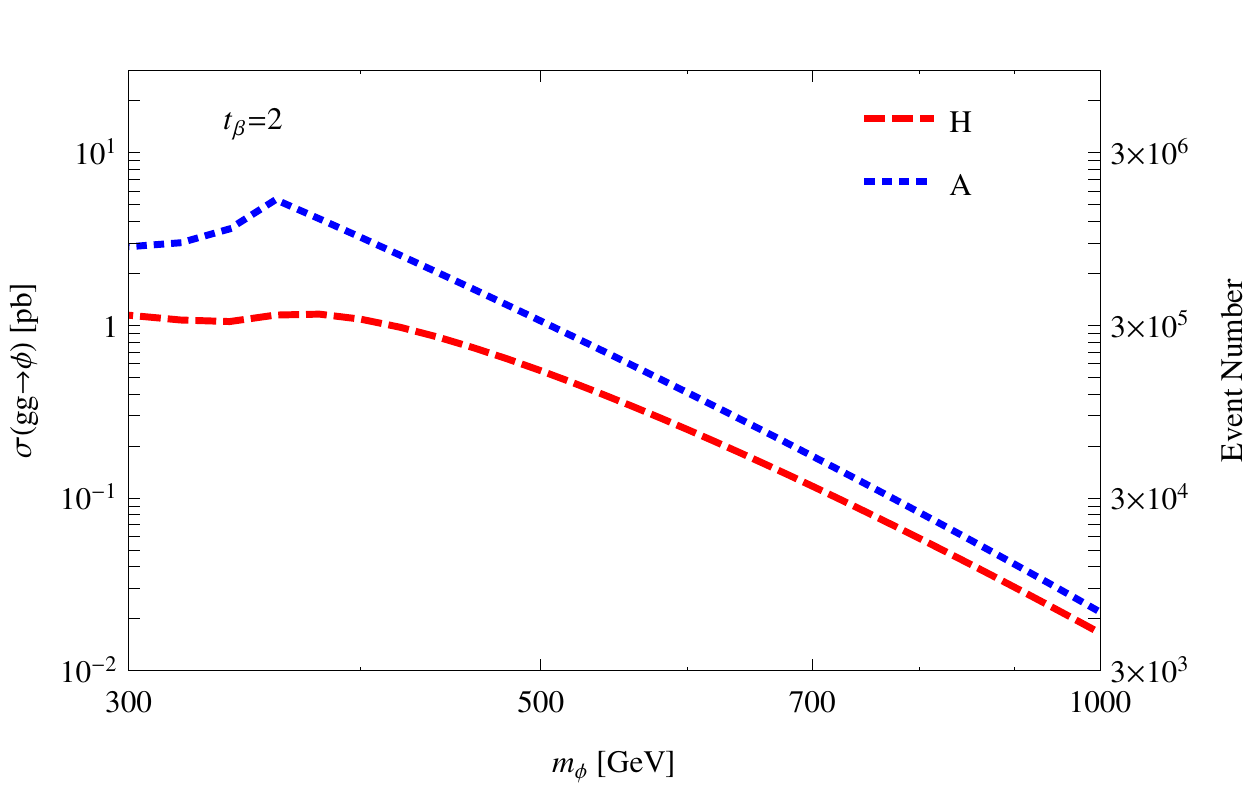}
  \caption{$CP$-even and $CP$-odd scalar boson production cross section via gluon fusion  as a function of the scalar boson mass at the LHC at $\sqrt{s}=$ 14 TeV in type-II THDM. We use $\sin(\alpha-\beta)\simeq 0.999$ and $t_\beta=2$. The right axis shows the annual event number achieved with an integrated luminosity of 300 fb$^{-1}$.}\label{GluonFusion}
\end{figure}

\subsection{$h\to Z\gamma\gamma$ branching ratio}
We now briefly discuss the  lightest $CP$-even scalar boson decay  $h\to Z\gamma\gamma$. Since $h$  must mimic  the properties of the SM Higgs boson,  it is expected that the $h\to Z\gamma\gamma$ branching ratio does not deviate considerably from its SM value. For $\sin(\beta-\alpha)\simeq 1$, the only contributions arise from box diagrams and the reducible diagram mediated by the $Z$ gauge boson. As mentioned before, the $hZA$  vertex  is considerably suppressed, whereas the $hHZ$ one is forbidden due to $CP$ invariance. Even more, there is no enhancement due to the $Z$-mediated reducible diagram since $m_h< 2m_Z$. For $m_h=125$ GeV and $t_\beta=10$ GeV we obtain $BR(h\to Z\gamma\gamma)\simeq 10^{-9}$, which does not deviate significantly from the SM value \cite{Abbasabadi:2008zz}. Therefore the new physics effects provided by the THDM does not give a significant enhancement to this decay and seem very far from the reach of detection.

\subsection{Kinematic distributions}
In the scenario in which the intermediary scalar boson is off-shell, the analysis of the behavior of some kinematic distributions could be helpful to disentangle the decay signal from the potential background. The $Z$ gauge boson energy distribution $d\Gamma(\phi_i \to Z\gamma\gamma)/dE_Z$ and   the photon invariant mass distribution $d\Gamma(\phi_i \to Z\gamma\gamma)/dm_{m_{\gamma\gamma^\prime}}$ could be useful for this task. To obtain the former,  one can plug the relation $dx_1=(2/m_{\phi_i})dE_Z$ into Eq. \eqref{DecayWidth} to obtain

\begin{equation}\label{ecuacionenerZA}
\dfrac{d\Gamma(\phi_i\to Z\gamma\gamma)}{dE_Z}=\dfrac{1}{128\pi^3}\int_{x_{2i}}^{x_{2f}}|\overline{\mathcal{M}}(\phi_i\to Z\gamma\gamma)|^2dx_2,
\end{equation}
where the $Z$ gauge boson energy is defined in the interval $(m_Z,(m_{\phi_i}^2+m_Z^2)/(2m_{\phi_i}))$.

 On the other hand, the expression for the photon invariant  mass distribution $d\Gamma(\phi_i \to Z\gamma\gamma)/dm_{m_{\gamma\gamma^\prime}}$ is obtained using the relation $dm_{\gamma\gamma'}=dE_Z/\sqrt{\mu_Z-x_1+1}$, which leads to

\begin{equation}\label{ecuacionmasinvA}
\dfrac{d\Gamma(\phi_i \to Z\gamma\gamma)}{dm_{\gamma\gamma^\prime}}=\dfrac{\sqrt{\mu_Z-x_1+1}}{128\pi^3}\int_{x_{2i}}^{x_{2f}}|\overline{\mathcal{M}}(\phi_i\to Z\gamma\gamma)|^2dx_2,
\end{equation}
where $m_{\gamma\gamma^\prime}$ is defined in the interval $( 0, m_{\phi_i}-m_Z)$.

\begin{figure}[ht!]
\includegraphics[width=8.5cm]{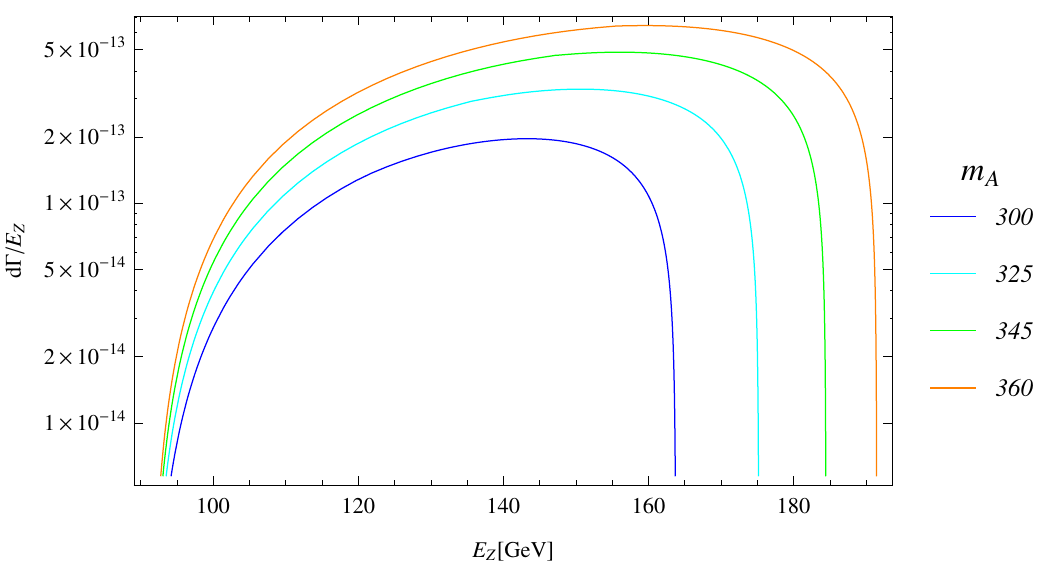}\hspace{0.2cm}\includegraphics[width=8.5cm]{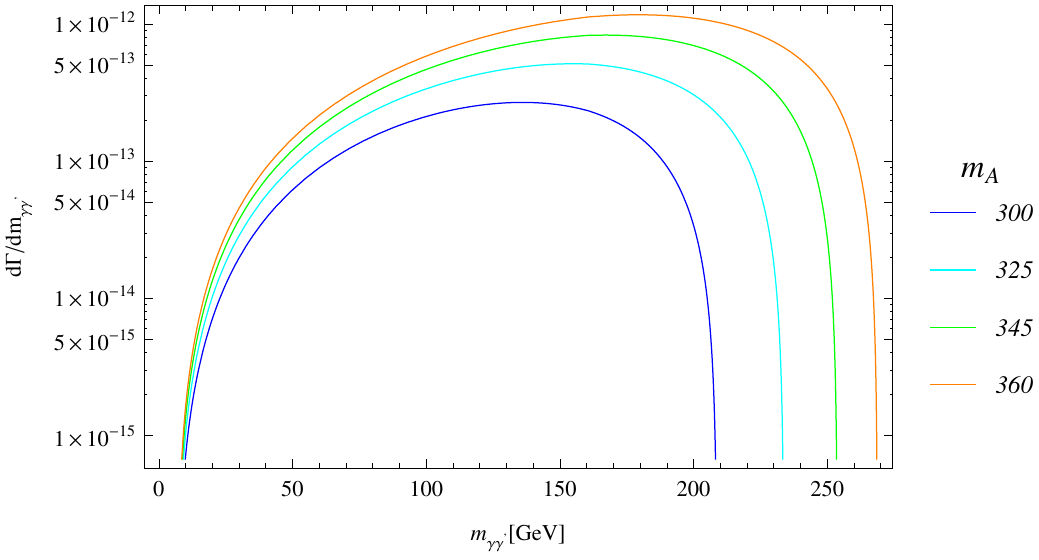}
     \caption{Energy (left plot) and photon invariant mass (right plot) distributions  $\Gamma(A\to Z\gamma\gamma)/dE_Z$ and $d\Gamma(A\to Z\gamma\gamma)/dm_{\gamma\gamma^\prime}$, for several masses of the $CP$-odd scalar boson in type-II THDM. We use  $m_H=700$ GeV and $t_\beta=15$. }
     \label{ADistributions}
\end{figure}

For illustrative purpose we show in Fig. \ref{ADistributions} the energy distribution $d\Gamma(A\to Z\gamma\gamma)/dE_Z$ and  the photon invariant mass distribution $d\Gamma(A\to Z\gamma\gamma)/dm_{\gamma\gamma} $ in the alignment limit for $m_H=700$ GeV, $m_{H^\pm}=750$ GeV, $t_\beta=15$, and a few  values of $m_A$. We observe that in the rest frame of the $CP$-odd scalar,  the $Z$ gauge boson   energy is peaked at about one half of $m_A$. A similar situation is observed for the invariant mass $m_{\gamma\gamma} $.

\subsection{$A\to Z\gamma\gamma$ and $H \to Z\gamma\gamma$ decays in type-I THDM}
\begin{figure}[ht!]
 \includegraphics[width=0.9\textwidth]{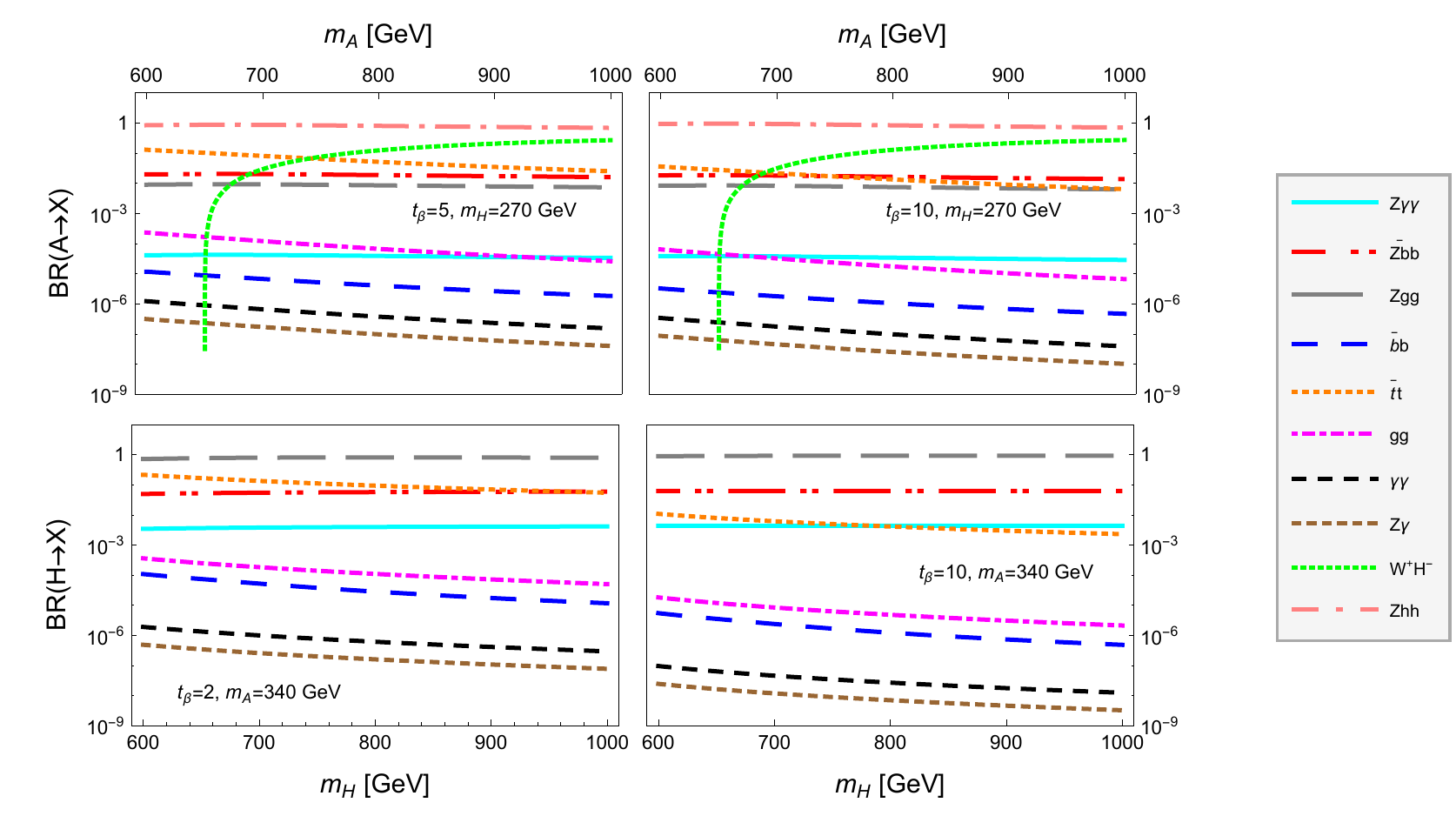}
\caption{Branching ratios for the $A\to Z\gamma\gamma$ and $H \to Z\gamma\gamma$ decays in type-I THDM as a function of the scalar boson masses for $m_{H^\pm}=570$ GeV, $\sin(\beta-\alpha)=0.999$, and two values of $t_\beta$ allowed by theory and experimental constraints.    The branching ratios for the main decay channels are also shown.\label{BRHtoZgg-type-I}}
\end{figure}

We now briefly analyze these decays in the framework of type-I THDM, where the charged scalar boson mass has a lower bound. Since the main contribution to the decays $A\to \gamma\gamma$ and $H\to \gamma\gamma$ arises from the top quark, the effect of a charged Higgs scalar boson with a mass less than $570$ GeV would not have a considerably effect on the decays we are interested in.  However, in type-I THDM the couplings of the $CP$-odd scalar boson to both quark types   are now proportional to $\cot\beta$, and the same is true for the couplings of $CP$-even  scalar bosons  in the $\cos(\alpha-\beta)\to 0$ limit \footnote{For the Feynman rules for type-I THDM see Ref. \cite{Branco:2011iw}.}. Therefore, the decay widths of the scalar bosons into  the $\bar{b}b$ pair would be suppressed for large $t_\beta$, which can do have an effect on our decays indeed. Consider for instance the decay $A\to Z\gamma\gamma$ in the scenario where $m_A> m_H+m_Z$. In type-II THDM the main decay channel is $A\to ZH\to Z\bar{b}b$, but for large $t_\beta$ this decay would get suppressed in type-I THDM as the $H\to \bar{b}b$ decay gets suppressed. This can translate into an enhancement of  the  $A\to ZH\to Z\gamma\gamma$ decay width. To analyze this scenario we have performed the explicit  calculation of the  $A\to Z\gamma\gamma$ and  $H\to Z\gamma\gamma$ decay widths in type-I THDM in the same scenarios considered in type-II THDM. The results for the respective branching ratios and those of the main decay channels are shown in Fig. \ref{BRHtoZgg-type-I}, where we observe that the $A\to Z\gamma\gamma$ and   $H \to Z\gamma\gamma$ decays can have an enhancement of about one order of magnitude with respect to the values obtained on type-II THDM.

\section{Conclusions}
\label{Conclusions}

In this work we have calculated the one-loop contributions to the decays  of the $CP$-odd and $CP$-even scalar bosons $A\to Z\gamma\gamma$ and $\phi\to Z\gamma\gamma$ ($\phi=h,H$)  in the framework of THDMs. We have presented analytical expressions for both box and reducible diagrams in terms of Passarino-Veltman scalar functions, though the main contributions arise from reducible diagrams. We first discuss the $A\to Z\gamma\gamma$ decay, which has not been discussed previously in the literature to our knowledge. For the numerical analysis we worked within the type-II THDM and considered a region of the parameter space still consistent with experimental data, with $\sin(\beta-\alpha)\simeq 1$, where the lightest $CP$-even scalar boson $h$ is identified with the SM Higgs boson, the $hZA$ vertex has a negligibly small strenght, and the heavy $CP$-even scalar does not couple to the weak gauge bosons. It was found that the $A\to Z\gamma\gamma$ branching ratio is only relevant in the scenario where $m_A>m_H+m_Z$, when the intermediary $H$ boson is on-shell.  For $m_A>600$ GeV and $t_\beta$ close to 1, $BR(A\to Z\gamma\gamma)$ can reach values of the order of $10^{-5}-10^{-4}$, but it decreases by about one order of magnitude as $t_\beta$ increases up to 10, which stems from the fact that the dominant contribution arises from the loops with the top quark, which couples to the scalar boson with a strength proportional to $1/t_\beta$.  On the other hand, when $m_A<m_H+m_Z$, $BR(A\to Z\gamma\gamma)$ is negligibly small, of the order of $10^{-10}$. As far as the $H\to Z\gamma\gamma$ decay is concerned, it exhibits a similar behavior and its branching ratio is non-negligible only in the scenario where  $m_H>m_A+m_Z$, when the $CP$-odd scalar is now on-shell. In this region of the parameter space, $BR(H\to Z\gamma\gamma)$ can reach the level of $10^{-4}-10^{-3}$ for $m_H>600$ GeV and $t_\beta\simeq 1$, but it decreases  for larger $t_\beta$. We also discussed the $h\to Z\gamma\gamma$ decay, which receives contribution from box diagrams and a reducible diagram mediated by the $Z$ gauge boson. Since the properties of the $h$ scalar boson are nearly identical to the SM Higgs boson, it is found that the $h\to Z\gamma\gamma$ branching ratio does not deviates significantly from the SM prediction and it is of the order of $10^{-9}$. Our calculation  is in agreement with previous evaluations.  The new physics effects of  THDMs are thus not relevant for this decay. Finally we also estimated these rare decays in the framework of type-I THDM, where we find that the respective branching ratios can be enhanced buy about one order of magnitude with respect to those of type-II THDM.

\acknowledgments{We acknowledge support from Consejo Nacional de Ciencia y Tecnolog\'ia and Sistema Nacional de Investigadores. Partial support from Vicerrector\'ia de Investigaci\'on y Estudios de Posgrado de la Ben\'emerita Universidad Aut\'onoma de Puebla is also acknowledged. }

\appendix

\section{Feynman rules}
\label{FeynmanRules}

\begin{figure}[ht!]
 \includegraphics[width= 12cm]{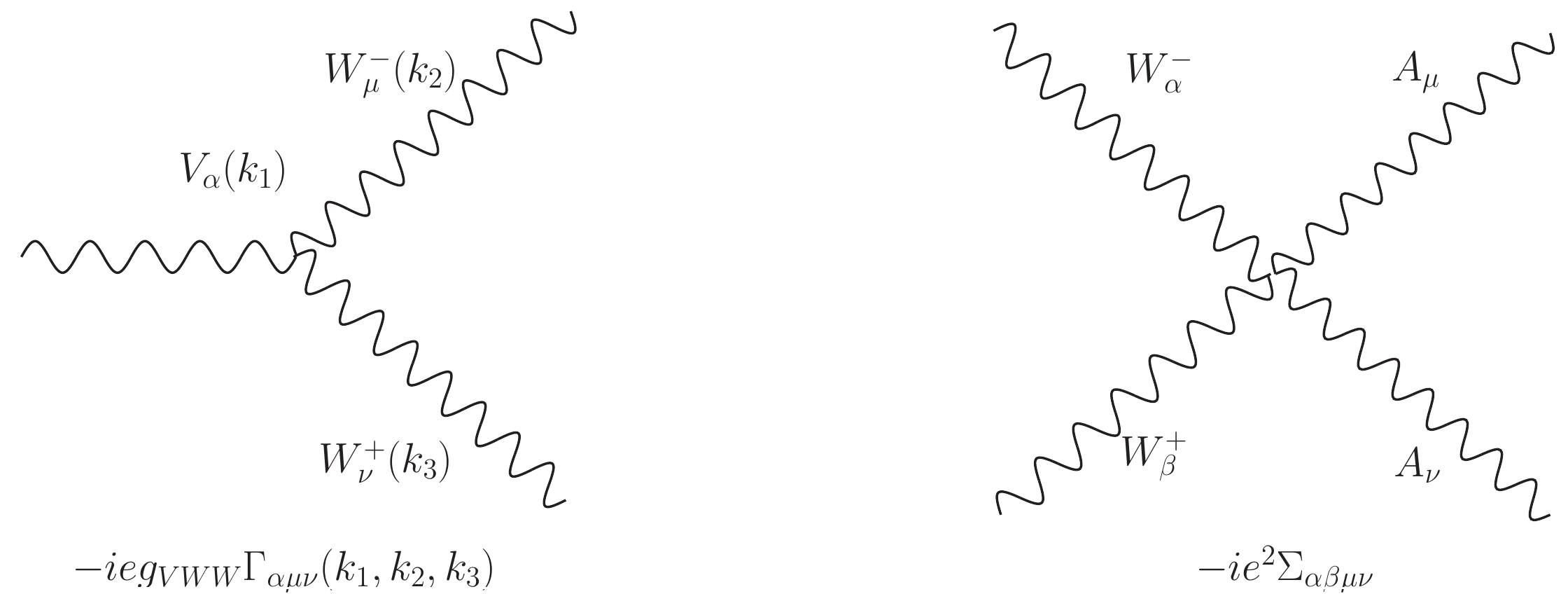}
 \caption{SM Feynman rules necessary for our calculation. All the 4-momenta are incoming. $g_{VWW}=1$ ($-\frac{c_W}{s_W}$) for $V=\gamma$ $(Z)$. In addition $\Gamma^{\alpha\mu\nu}(k_1,k_2,k_3)=(k_1-k_2)^\nu g^{\alpha\mu}+(k_2-k_3)^\alpha g^{\mu\nu}+(k_3-k_1)^\mu g^{\alpha\nu}$ and $\Sigma^{\alpha\beta\mu\nu}=2g^{\alpha\beta}g^{\mu\nu}-g^{\alpha\mu}g^{\beta\nu}-g^{\alpha\nu}g^{\beta\mu}$. We also need the  Feynman rules for the interactions of the photon and the $Z$ gauge boson with a fermion pair, which are as follows: $-ieQ_f \gamma^\mu$  and $-i\frac{g}{2c_W}(g_V^f-g_A^f\gamma^5) \gamma^\mu$, respectively, where $g_A^f=\frac{1}{2}T^3_f$ and  $g_V^f=\frac{1}{2}T_3^f-Q_f s_W^2$, with $Q_f$ the fermion charge and $T^3_f=1$ ($-1$) for up quarks (down quarks and charged leptons) .\ \label{SMFeynmanRules}}
\end{figure}

In this appendix we present the Feynman rules necessary for our calculation, which was performed in the unitary gauge. We first present  the Feynman rules for the vertices $V W^-W^+$, $\gamma\gamma W^-W^+$, and $V\bar{f}f$ ($V=\gamma, Z$), which are identical to the SM ones and are shown in Fig. \ref{SMFeynmanRules}.

\begin{figure}[ht!]
 \includegraphics[width= 12cm]{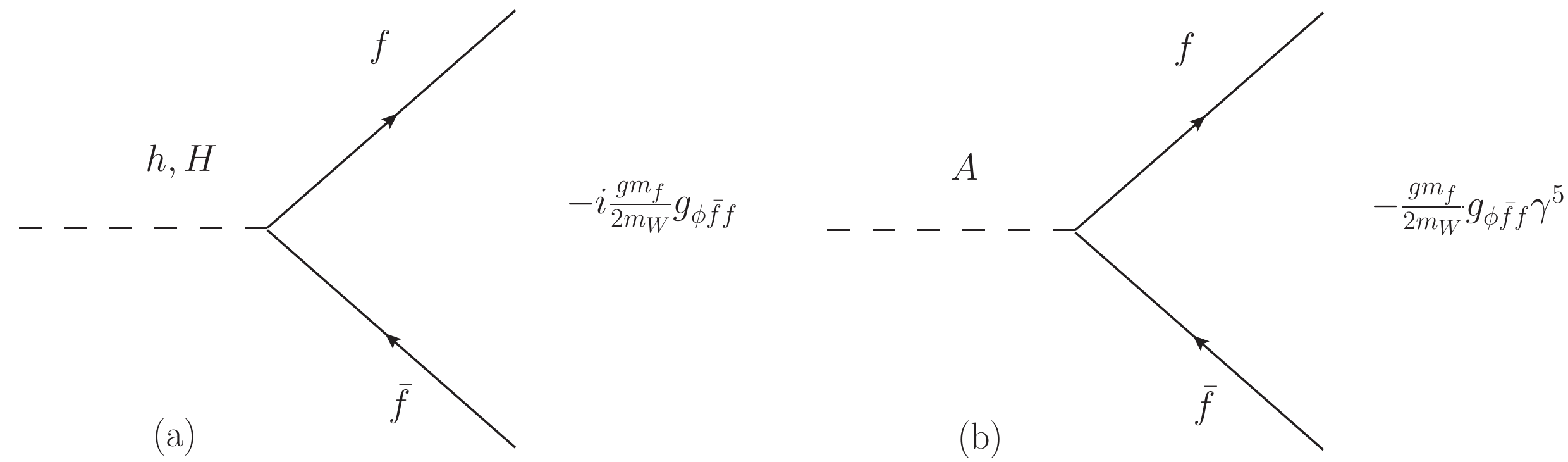}
 \caption{Feynman rules for the couplings of the scalar bosons to fermions in THDMs.  The corresponding coupling constants for type-II THDM are shown in Table \ref{CouplingConstants}. \label{FeynmanRules1}}
\end{figure}

We  also need  the couplings of the neutral Higgs bosons to the fermions, the gauge bosons, and the charged scalar bosons.
The Feynman rules for the couplings of the neutral Higgs bosons to  fermion pairs in THDMs are  shown in Fig. \ref{FeynmanRules1}, and the corresponding coupling constants for type-II THDM are presented and Table \ref{CouplingConstants} \cite{Branco:2011iw}.

\begin{table}[!hbt]
\begin{center}
\caption{Constants for the couplings of the scalar bosons to fermions and gauge bosons in type-II THDM as described in Figs. \ref{FeynmanRules1} and \ref{FeynmanRules2}.  We have used the short-hand notation $s_a=\sin a$ and $ c_a=\cos a$. The $g_{\phi ZZ}$ couplings obey $g_{\phi ZZ}=\frac{1}{c_W^2}g_{\phi W W}$ \cite{Branco:2011iw}.}
\label{CouplingConstants}
\begin{tabular}[t]{c c c c c c c}
\hline
\hline
$\phi$  &$g_{\phi uu}$ 	 &$g_{\phi dd}$ ($g_{\phi ll}$)
&$g_{\phi WW}$ &$g_{\phi ZA}$ &$g_{\phi H^-H^+}$&$g_{\phi W^-H^+}$\\
\hline \hline
$h$ &$-\left(s_{\beta-\alpha}+\dfrac{c_{\beta-\alpha}}{ t_\beta}\right)$  &
$-\left(s_{\beta-\alpha}-t_\beta c_{\beta-\alpha}\right)$ &
$s_{\beta-\alpha}$ &$c_{\beta-\alpha}$ &$\left(c_Wc_{\beta-\alpha}-\frac{1}{2c_W}c_{2\beta} c_{\beta+\alpha}\right)$&
$c_W s_{\beta-\alpha}$\\
\hline
$ H$ &$-\left(c_{\beta-\alpha}-\dfrac{s_{\beta-\alpha}}{ t_\beta}\right)$  &$-\left(c_{\beta-\alpha}+t_\beta s_{\beta-\alpha}\right)$ &$c_{\beta-\alpha}$ &$-s_{\beta-\alpha}$ &$ \left( c_W s_{\beta-\alpha}+\frac{1}{2c_W}c_{2\beta} s_{\beta+\alpha}\right) $&
$c_W c_{\beta-\alpha}$\\
\hline
$A$  &$\dfrac{1}{t_\beta}$  &$t_\beta$ &$0$	&$0$ &$0$&$-ic_W$\\
\hline
\end{tabular}
\end{center}
\end{table}

\begin{figure}[ht!]
  \includegraphics[width= 12cm]{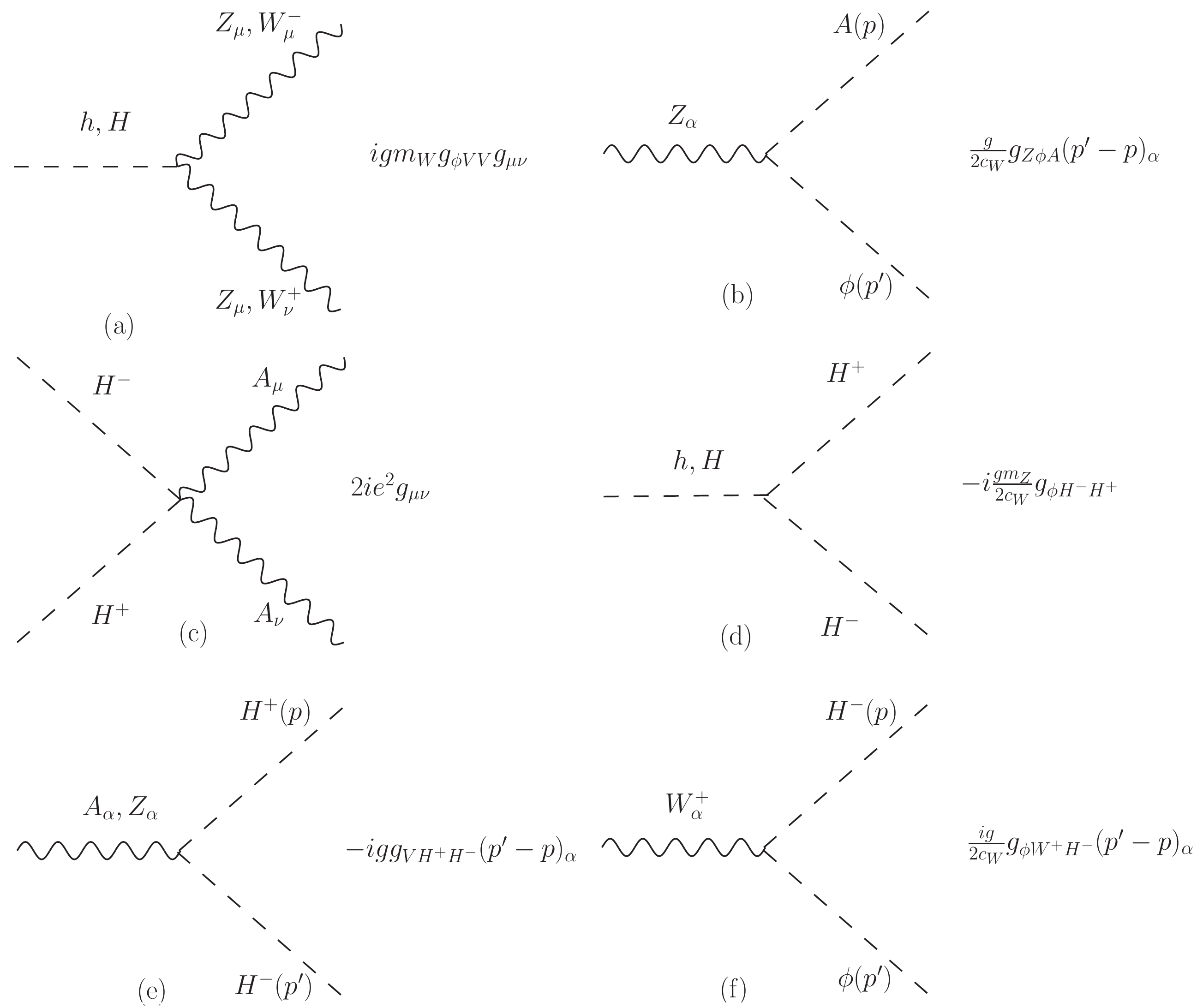}
 \caption{Feynman rules necessary for our calculation in THDMs. All the four-momenta are incoming.  Here $\phi=h,H$ in diagram (b) but $\phi=h,H,A$ in diagram (f).  Also, in diagram (e) $g_{AH^+H^-}=s_W$ and $g_{ZH^+H^-}=c_{2W}/{2c_W}$. The remaining coupling constants for type-II THDM are presented in Table \ref{CouplingConstants}. \label{FeynmanRules2}}
\end{figure}

As far as the couplings of  scalar bosons to  gauge bosons,  we must expand the covariant derivative  of Eq. \eqref{Lag} in terms of the physical fields. It is straightforward to obtain the Feynman rules shown in Fig. \ref{FeynmanRules2}  for the couplings $\phi VV$ ($V=W,Z$)  and $Z\phi A$  ($\phi=h,H$). Note that the $AVV$ ($V=W,Z$) and $HhZ$ couplings are absent due to $CP$ conservation.    Other Feynman rules such as those for the vertices $\gamma H^-H^+$, $Z H^-H^+$,  and $\gamma\gamma H^-H^+$ are also obtained from the Higgs kinetic sector and are  shown in Fig. \ref{FeynmanRules2}, and so is the Feynman rule for the couplings of the $CP$-even scalar bosons to a pair of charged scalar boson $ \phi H^-H^+$, which emerge from the Higgs potential \eqref{pot} once it is diagonalized.

\section{$A\to Z\gamma\gamma$ and $\phi\to Z\gamma\gamma$ ($\phi=h,H)$ decay amplitudes}
\label{Amplitudes}
We now present the form factors of Eqs. \eqref{AtoZggAmplitude} and \eqref{phitoZggAmplitude} in terms of Passarino-Veltman scalar functions.

\subsection{$A\to Z\gamma\gamma$ decay}

\subsubsection{Box diagrams}

Box diagrams give the following contributions to the  form factors of Eq. \eqref{AtoZggAmplitude}
\begin{equation}
\mathcal{F}_{i}^{Box}=\sum_f\frac{16g^f_A g_{A\bar{f}f} g^2 \alpha m_f^2 Q_f^2N_c^f}{ m_W c_W s X_A ^2} f_i^{Box},
\end{equation}
with
\begin{equation}\label{FFB1}
\begin{split}
f_{1}^{Box}&= s\Bigl[X_A \Delta_{2Z}+sm_Z^2(s_1+s_2)\Bigr]C_1(s)+{\frac{\Delta_{2A}}{ s}}\Delta_{2Z}\Bigl[s_2 \Delta_{1Z}^2-m_Z^2\Delta_{2A}^2\Bigr]C_2(s_2)+{\frac{\Delta_{1A}}{ s}}\Bigl[s_1 (s_1+s)\Delta_{2Z}^2\\
&+[m_A^2(2s_1+s)(s_1-\Delta_{2Z})+(m_Z\Delta_{1A})^2-(s_1 s+(s_1+s)^2)s_1]m_Z^2 \Bigr]C_2(s_1)-\Delta_{1Z}^2\Bigl[\Delta_{2Z}\Bigl(m_Z^2-{\frac{X_A}{ s}}\Bigr)\\
&+sm_Z^2\Bigr]C_3(s_1)+\Delta_{2A}\Delta_{2Z}^2\Bigl[m_Z^2-{\frac{X_A}{ s}}\Bigr]C_3(s_2)+sm_Z^2 \Bigl[(s_1-s_2)^2+X_A \Bigr]C_4(s)+{\frac{s}{ 2}}\Bigl[m_A^2m_Z^2(s_1(\Delta_{1Z}\\
&+s_1-s_2)-2X_A )+s_1^2(2 m_Z^4-(2 s_1+s_2) m_Z^2+s_2^2)+\Bigl(m_Z^2-{\frac{X_A}{ s}}\Bigl)4m_f^2X_A \Bigr]D_1(s_1)-{\frac{s}{2}} \Bigl[s_1 s_2^3\\
&-s_2m_Z^2(s_2(\Delta_{2Z}+s_1+3s)+X_A )-4Xm_f^2\Bigl(m_Z^2-{\frac{X_A}{ s}}\Bigr)\Bigr]D_1(s_2)+{\frac{1}{ 2}} \Bigl[s_2s^2(s_2-4 m_f^2)(m_Z^2+s_2)\\
&+\Delta_{2A}\Delta_{2Z}\Bigl({\frac{2\Delta_{2A}^2}{ s}}\Delta_{2Z}^2-\Delta_{2A}(4 m_f^2-m_Z^2-5 s_2)\Delta_{2Z}+2(s_2-2m_f^2)(m_Z^2+2s_2)s\Bigr) \Bigr]D_2(s_2),
\end{split}
\end{equation}

\begin{equation}
\begin{split}
f_{2}^{Box}&=\frac{X_A}{2} \Bigl(2\Delta_{2A}\Bigl(sC_1(s) +\Delta_{1A}C_2(s_1)+{\frac{\Delta_{2Z}}{ 2}}C_3(s_2)\Bigr)+ sm_A^2\Delta_{1Z}D_1(s_1)\\
&+\Delta_{1Z} X_A D_2(s_2)- s\Big[X_A +s_2\Delta_{2A}\Big]D_1(s_2)\Bigr),
\end{split}
\end{equation}

and
\begin{equation}\label{FFB3}
\begin{split}
f_{3}^{Box}&=2sm_A^2\Big[X_A +s(s_1+s_2)\Big]C_1(s)+2ss_1 \Delta_{1A}C_2(s_1) -2\Delta_{2A}^2\Delta_{2Z}C_2(s_2)\\
&-2\Delta_{1Z}^2\Delta_{1A}C_3(s_1)+2ss_2\Delta_{2Z}C_3(s_2)-2s\Bigl[X_A +(s_1-s_2)^2\Bigr]C_4 (s)-X_A \Bigl[X_A+4s m_f^2\Bigr]D_2(s_2)\\
&-s\Big[s_1 (X_A +2s_1s)+4X_A  m_f^2\Big]D_1 (s_1)- s\Big[s_2(X_A +2s_2s)+4 X_A m_f^2
 \Bigr]D_1(s_2),
 \end{split}
\end{equation}
where the kinematical invariant variables $ s_1$, $s_2$ and $s$ were defined in Eqs. \eqref{s1}-\eqref{s}. In addition, we use the following auxiliary variables:

\begin{subequations}
\begin{eqnarray}
\Delta_{ij}&=&s_i-m_j^2,\\
 X_\chi&=&s_1s_2-m_Z^2m_\chi^2,
\end{eqnarray}
\end{subequations}
for $i=1,2$ and $j=A,Z$.
As for the  three- and four-point
Passarino-Veltman scalar functions $C_i$ and $D_i$, they  are defined as

\begin{equation}
\begin{split}
C_1(p^2) &=C_0(0, 0, p^2, m_f^2, m_f^2, m_f^2),\\
C_2(p^2)&=C_0(0,p^2, m_A^2 , m_f^2, m_f^2, m_f^2),\\
C_3(p^2)&=C_0(m_Z^2, 0,p^2, m_f^2, m_f^2, m_f^2),\\
C_4(p^2) &= C_0(m_Z^2, p^2, m_A^2 , m_f^2, m_f^2, m_f^2),\\
D_1(p^2) &= D_0(m_Z^2, 0, 0, m_A^2 , p^2, s,m_f^2, m_f^2, m_f^2, m_f^2),\\
D_2(p^2) &= D_0(m_Z^2, 0, m_A^2, 0,s_1, p^2, m_f^2, m_f^2, m_f^2, m_f^2).
\end{split}
\end{equation}

As we can see for Eq. \eqref{FFB1}-\eqref{FFB3}  the box diagrams  amplitudes are free of ultraviolet divergences since are free of two-point Passarino-Veltman scalar functions.

\subsubsection{Reducible diagram contribution}
The reducible diagram of Fig. \ref{FeynmanDiagramsAtoZggRed} only contribute to the form factor $\mathcal{F}_1$ of Eq. \eqref{AtoZggAmplitude}. The  Passarino-Veltman technique allowed us to obtain the following results for the fermion and $W$ gauge boson contributions

\begin{equation}
\mathcal{F}_{1}^\chi=\dfrac{2g^2 \alpha g_{\phi ZA}} {c_W\; s(m_\phi^2-s)}\left\{
\begin{array}{llll}\sum_f
\dfrac{g_{\phi \bar{f}f} m_f^2 Q_f^2N_c^f}{m_W}\Bigl[1+\Bigl(2m_f^2- \dfrac{s}{2}\Bigr)C(s,m_\chi^2)\Bigr]&&\chi=f,\\
-\dfrac{g_{\phi WW}}{4 m_W}
\Bigl[ \dfrac{s}{2}+3m_W^2(1+(2m_W^2-s)C(s,m_W^2))\Bigr]&&\chi=W,\\
\dfrac{g_{\phi H^\pm H^\pm}m_Z}{4 }
\Bigr[2m_{H^\pm}^2C(s,m_{H^\pm}^2)
+1\Bigr]&&\chi=H^{\pm}.
\end{array}\right.,
\end{equation}
where the three-point scalar function $C(s,m_\chi^2)$ can be written in terms of elementary functions as follows
\begin{equation}
C(s,m_\chi^2) =C_0(0,0,s,m_\chi^2,m_\chi^2,m_\chi^2)=-\frac{2}{s} f\left(\frac{4m_\chi^2}{s}\right),
\end{equation}
where $f(x)$ is given in  Eq. \eqref{f(x)}.

\subsection{$\phi\to Z\gamma\gamma$ ($\phi=h,H$) decay}
\subsubsection{Box diagram contribution}

The box diagram contributions to the form factors of Eq. \eqref{phitoZggAmplitude} are given as follows

\begin{equation}
\mathcal{G}_{i}^{Box}=\sum_{\phi=h,H}\sum_f \dfrac{16g^f_Ag_{\phi\bar{f}f}  g^2 \alpha m_f^2 Q_f^2N_c^f}{m_W c_WX_{\phi}^2 }{g}_{i}^{Box},
\end{equation}
with
\begin{equation}
\begin{split}
g_{1}^{Box} &=\frac{X_{\phi}}{32}\Big(2 \Delta_{2{\phi} }C_2(s_2)-2 \Delta_{1{\phi} }C_2(s_1)+2
   (s_1-s_2) C_4(s)+\Delta_{2{\phi} }\Delta_{2Z}D_1(s_2)-\Delta_{1Z}\Delta_{1{\phi} }D_1(s_1)\Big),
\end{split}
\end{equation}

\begin{equation}
\begin{split}
g_{2}^{Box} &=m_\phi^2\Biggl(-2\Big[X_{\phi} +s(s_1+s_2)\Big]C_1(s) +\dfrac{2}{s}\Delta_{1Z}\Delta_{1{\phi} }^2 C_2(s_1)+2s_2\Delta_{2{\phi} }C_2(s_2)-2\Delta_{1Z}sC_3(s_1)-\dfrac{2}{s}\Delta_{2Z}^2\Delta_{2{\phi} }C_3(s_2)\\ &+2\Big[2X_{\phi} +(s_1-s_2)^2 \Big]C_4(s)+\Big[4m_f^2X_{\phi} +s_1(X_{\phi} +2s_1s) \Big]D_1(s_1)
-\Big[4m_f^2X_{\phi} +s_2(X_{\phi} +2s_2s) \Big]D_1(s_2) \\
&+\dfrac{1}{s}\Big[ X_{\phi} (X_{\phi} +4sm_f^2) \Big]D_2(s_2)\Biggr),
\end{split}
\end{equation}

\begin{equation}
\begin{split}
g_{3}^{Box} &=\dfrac{1}{s^2}\Big(-s^2\Big[s_1s_2^2+m_Z^2(m_{\phi} ^2(m_Z^2+2s_2)+\Delta_{2Z}s_2) \Big]C_1(s)-\Delta_{1{\phi} }\Big[ m_Z^8-(3 s+2(s_1+s_2)) m_Z^6\\
&+(2 s^2+3 s_1
   s+s_1^2+s_2^2+4 (s+s_1) s_2)
   m_Z^4-((s_1+s_2) s^2+s_2 (4
   s_1+s_2) s+2 s_1 s_2 (s_1+s_2))m_Z^2\\
   &+s_1 (s+s_1) s_2^2\Big]C_2(s_1)+\Delta_{2{\phi} }\Delta_{2Z}\Big[ m_Z^2\Delta_{2{\phi} }^2-\Delta_{1Z}^2s_2 \Big]C_2(s_2)+\Delta_{2{\phi} }\Delta_{2Z}^2\Big[X_{\phi} -sm_Z^2 \Big]C_3(s_2)\\
   &+\Delta_{1Z}^2\Big[sm_Z^2(s+2s_2-m_Z^2) -\Delta_{1Z}\Delta_{2Z}^2 \Big]C_3(s_1)-s^2m_Z^2\Big[ 2X_{\phi} +(s_1-s_2)^2  \Big]C_4(s)+\frac{1}{2} s \Big[4 X_\phi (X_\phi\\
   &-sm_Z^2) m_f^2+s (-2
   m_Z^8+(4 s+3 s_1+4 s_2) m_Z^6-(2 s^2+3
   s_1 s+4 s_2 s+s_1^2+2 s_2^2+6 s_1 s_2) m_Z^4\\
   &+s_1
   (3 s_2^2+3 s s_2+2 s_1 s_2-2 s s_1)
   m_Z^2-s_1^2 s_2^2)\Big] D_1(s_1)+
   \frac{1}{2} s \Big[4 X_\phi (X_\phi -sm_Z^2) m_f^2+s s_2
   (m_Z^4(s_2+m_\phi^2)\\
   &-s_2 (3
   s+2 s_1+s_2) m_Z^2+s_1 s_2^2)\Big]D_1(s_2) +\frac{1}{2} X_\phi \Big[2 m_Z^8-(5 s+4
   (s_1+s_2)) m_Z^6+(3 s^2+4 m_f^2
   s+5 s_1 s\\
   &+6 s_2 s+2 s_1^2+2 s_2^2+8 s_1 s_2)
   m_Z^4-(4 s (2 s+s_1+s_2) m_f^2+s_2
   (s^2+(6 s_1+s_2) s+4 s_1
   (s_1+s_2))) m_Z^2\\
   &+s_1 s_2 (4
   s m_f^2+(s+2 s_1) s_2)\Big] D_2(s_2)\Big),
\end{split}
\end{equation}
and
\begin{equation}
\begin{split}
g_4^{Box} &=\dfrac{1}{8\Delta_{1Z}\Delta_{2Z}}\Big( 4m_Z^2X_{\phi}[ \Delta_{1Z}\Delta B(m_Z^2,s_2)+\Delta_{2Z}\Delta B(m_Z^2,s_1)]+4\Delta_{1Z}\Delta_{2Z}X_{\phi } \Delta B(s,m_A^2)\\
&+2 s\Delta_{1Z}\Delta_{2Z}\Big[2X_\phi+m_Z^2(m_Z^2-s)-s_1(s_2
   +s_1) +s_2^2\Big]C_1(s)+\Delta_{1Z}\Delta_{2Z}
   (s+\Delta_{2Z})
   \Big[2s_1\Delta_{1Z}\\
   &-X_\phi\Big]C_2(s_1)-\Delta_{1Z}\Delta_{2Z}(s+\Delta_{1Z})
   \Big[5X_\phi-2s_2(\Delta_{1Z}+s_1)+2s_2^2 \Big]C_2(s_2)
   -\Delta_{1Z}^2\Delta_{2Z}
   \Big[2s_1\Delta_{1Z}\\
   &-X_\phi\Big]C_3(s_1)+\Delta_{1Z}\Delta_{2Z}^2
   \Big[5X_\phi-2s_2(\Delta_{1Z}+s_1)+2s_2^2 \Big]C_3(s_2)
   +2 \Delta_{1Z}  \Delta_{2Z}\Big[2m_Z^4(m_\phi^2+2(s_1+s_2))\\
&-(5 s_1^2-3
   s_2^2+4 s (s_1-s_2))
   m_Z^2+(s_1-s_2) (s_1+s_2)^2\Big]C_4(s)+\Delta_{1Z}\Delta_{2Z}\Big[-2m_Z^6(m_Z^2+2m_\phi^2)\\
&+(2 s^2+(5
   s_1+4 s_2) s+2 (s_1^2+4 s_2
   s_1+s_2^2)) m_Z^4-s_1 (s^2-(s_1-5
   s_2) s+4 s_2 (s_1+s_2))
   m_Z^2+s_1^2 (2 s_2^2\\
&+s (s_2-2
   s_1))-4 m_f^2 \Delta_{1Z}
   X_\phi\Big]D_1(s_1)-\Delta_{1Z}\Delta_{2Z}\Big[-2m_Z^6(m_Z^2+2m_\phi^2)+(2
   (s+s_1)^2+2 s_2^2\\
&+(9 s+8 s_1)
   s_2) m_Z^4-s_2 (5 s^2+3 (3
   s_1+s_2) s+4 s_1 (s_1+s_2))
   m_Z^2+s_2^2 (s_1 (s+2 s_1)+2 s
   s_2)+4 m_f^2 (m_Z^2\\
&-2 s_1+s_2)
  X_\phi\Big]D_1(s_2)+4 m_f^2 \Delta_{1Z}\Delta_{2Z}
   (s_1-s_2)X_{\phi} D_2(s_2)
\Big),
\end{split}
\end{equation}
with the two-point Passarino-Veltman scalar functions defined as $\Delta B(r_1^2,r_2^2)=B_0(r_1^2,m_f^2,m_f^2)-B_0(r_2^2,m_f^2,m_f^2)$.  It is also evident that ultraviolet divergence cancel out.

\subsubsection{Reducible diagram contribution}
The reducible diagrams related to the processes $\phi\to Z\chi^*\to Z\gamma\gamma$, with $\chi=A,Z$, yield the following contribution to the  form factor of Eq. \eqref{GRD}

\begin{equation}
\mathcal{G}_{3}^{\chi}=\sum_f \dfrac{  g^2 \alpha Q_f^2 m_f^2N_c^f}{ 2c_W m_Z\pi}\left\{\begin{array}{lcr}
-\dfrac{g_{A\bar{f}f}g_{\phi Z A}}{ c_W(m_A^2-s)}\;C(s,m_f^2)&&\chi=A,\\
\dfrac{2g^f_A g_{\phi ZZ} }{s}\;C(s,m_f^2)&&\chi=Z.
\end{array}\right.
\end{equation}

\section{Squared average amplitudes}
\label{SquaredAmplitudes}

From the general form of the invariant amplitudes for the $A \to Z\gamma\gamma$ and $\phi\to Z\gamma\gamma$ ($\phi=h,H$) decays presented in Eqs. \eqref{AtoZggAmplitude} and \eqref{phitoZggAmplitude}, respectively, we can readily obtain the square amplitudes averaged over photon and $Z$ polarizations, which are required for the calculation of the decay width \eqref{DecayWidth}. The  results can be written as follows.
\subsection{ $A \to Z\gamma\gamma$ decay}

\begin{align}
\label{M2ADecay}
|\overline{\mathcal{M}}(A\to Z\gamma\gamma)|^2&=\frac{m_A^6}{4}\Biggl(
\frac{\hat{s}^2\hat{\Delta}_{1Z}^2}{2 \mu _Z} \left| \mathcal{F}_1\right| ^2+
\frac{1}{2 \mu _Z}\zeta_3\left| \mathcal{F}_2\right| ^2+
\frac{1}{8 \mu _Z} \hat{\Delta}_{2Z}^2 \zeta_2\left| \mathcal{F}_3\right| ^2+
\frac{\hat{s}^2\zeta_1}{2 \mu _Z} {\rm Re}\left[ \mathcal{F}_1 \,\tilde{\mathcal{F}}_1^*\right]\nonumber\\&-
\frac{1}{2 \mu _Z} \hat{\Delta}_{1Z} \hat{\Delta}_{2Z}  \left( 2\hat{s} \mu _Z- \hat{\Delta}_{1Z}\hat{\Delta}_{2Z}\right){\rm Re}\left[ \mathcal{F}_2 \,\tilde{\mathcal{F}}_2^*\right]+
\frac{1}{8 \mu _Z}\zeta_1\zeta_2 {\rm Re}\left[ \mathcal{F}_3 \,\tilde{\mathcal{F}}_3^*\right]\nonumber\\&+
 \hat{s}^2 \hat{\Delta}_{1Z}{\rm Re}\left[ \mathcal{F}_1 \mathcal{F}_2^*\right]+
\frac{1}{2} \hat{s}^2\zeta_1 \zeta_2 {\rm Re}\left[ \mathcal{F}_1 \mathcal{F}_3^*\right]+
\frac{1}{2 \mu _Z}\hat{\Delta}_{1Z} \hat{\Delta}_{2Z}^2  \zeta_1\zeta_2{\rm Re}\left[ \mathcal{F}_2 \mathcal{F}_3^*\right]\nonumber\\&-
 \hat{s}^2 \hat{\Delta}_{1Z}{\rm Re}\left[ \mathcal{F}_1 \,\tilde{\mathcal{F}}_2^*\right]+
\frac{1}{2} \hat{s}^2\hat{\Delta}_{1Z}^2 {\rm Re}\left[ \mathcal{F}_1 \,\tilde{\mathcal{F}}_3^*\right]+
\frac{1}{2 \mu _Z}\hat{\Delta}_{1Z} {\rm Re}\left[ \mathcal{F}_2 \,\tilde{\mathcal{F}}_3^*\right]\Biggl)+\left(\hat{s}_1\leftrightarrow\hat{s}_2\right),
\end{align}
where $\hat{s}_i=s_i/m_A^2$,  $\hat{s}=s/m_A^2$, $\hat{\Delta}_{ij}={\Delta}_{ij}/m_A^2$,  $\tilde{\mathcal{F}}_i(s,s_1,s_2)=\mathcal{F}_i(s,s_2,s_1)$. Also

\begin{equation}
\label{lambda1}
\zeta_1=\mu _Z^2-\left(2\hat s+s_1+\hat s_2\right) \mu _Z+\hat s_1 \hat s_2,
\end{equation}

\begin{equation}
\label{lambda2}
\zeta_2=\left(\mu _Z^4-2 \left(\hat{s}+\hat{s}_1+\hat{s}_2\right) \mu _Z^3+\left(2 \hat{s}^2+2 \left(\hat{s}_1+\hat{s}_2\right) \hat{s}+\hat{s}_1^2+\hat{s}_2^2+4 \hat{s}_1 \hat{s}_2\right) \mu _Z^2-2 \hat{s}_1 \hat{s}_2 \left(\hat{s}+\hat{s}_1+\hat{s}_2\right) \mu _Z+\hat{s}_1^2 \hat{s}_2^2\right),
\end{equation}
and
\begin{equation}
\label{lambda3}\zeta_3=\hat{s}_1^2 \hat{\Delta}_{2Z}^2+2 \hat{s}_1 \left(\hat{s}-\hat{\Delta}_{2Z}\right) \mu _Z \hat{\Delta}_{2Z}+\mu _Z^2 \left(-2 \hat{s}^2+2 \mu _Z \hat{s}+\hat{s}_2^2+\mu _Z^2-2 \hat{s}_2 \left(\hat{s}+\mu _Z\right)\right).
\end{equation}

\subsection{ $\phi \to Z\gamma\gamma$ ($\phi=h,H$) decay}
From Eq. \eqref{phitoZggAmplitude} we obtain
\begin{align}\label{M2phiDecay}
|\overline{\mathcal{M}}(\phi\to Z\gamma\gamma)|^2&=\frac{\hat s m_\phi^6}{2}\Bigl(
 \eta_2  \left| \mathcal{G}_1\right| ^2
-\frac{1}{4} \hat\Delta_{1Z}^2\eta_1 \left| \mathcal{G}_2\right| ^2
+\frac{\hat s \hat\Delta_{1Z}^2}{4 \mu _Z} \left| \mathcal{G}_3\right| ^2
+\frac{1}{4\hat s \mu _Z}\eta_3 \left| \mathcal{G}_4\right| ^2
-\eta_2  {\rm Re}\left[\mathcal{G}_1\tilde{\mathcal{G}}_1^*\right]
\nonumber\\&+ \hat\Delta _{2 Z}\eta_1 {\rm Re}\left[\mathcal{G}_1 \tilde{\mathcal{G}}_2^*\right]
-\hat s \hat\Delta _{2 Z}{\rm Re}\left[\mathcal{G}_1 \tilde{\mathcal{G}}_3^* \right]
- \eta_2  {\rm Re}\left[\mathcal{G}_1 \tilde{\mathcal{G}}_4^*\right]
+\frac{1}{8} \hat\Delta_{1Z} \hat\Delta _{2 Z} \eta_1 {\rm Re}\left[\mathcal{G}_2\tilde{\mathcal{G}}_2^*\right]
+\frac{1}{2} \hat s\eta_1 {\rm Re}\left[ \mathcal{G}_2 \tilde{\mathcal{G}}_3^*\right]
\nonumber\\&+ \hat\Delta_{1Z}
 \eta_1\frac{1}{2} \hat s {\rm Re}\left[\mathcal{G}_2 \tilde{\mathcal{G}}_4^*\right]
+\frac{1}{2 \mu _Z}  \left[\hat\Delta_{1Z} \hat\Delta _{2 Z}-2 \hat s \mu _Z\right]{\rm Re}\left[ \mathcal{G}_3 \tilde{\mathcal{G}}_3^*\right]
- \hat\Delta_{1Z}{\rm Re}\left[\mathcal{G}_3 \tilde{\mathcal{G}}_4^*\right]
\nonumber\\&
-\frac{1}{4 \hat s\mu _Z}\eta_3 {\rm Re}\left[\mathcal{G}_4 \tilde{\mathcal{G}}_4^*\right]
- \hat\Delta_{1Z}
 \eta_1{\rm Re}\left[\mathcal{G}_1 \mathcal{G}_2^*\right]+\hat s \hat\Delta_{1Z}{\rm Re}\left[\mathcal{G}_1 \mathcal{G}_3^*\right]
+\eta_2  {\rm Re}\left[\mathcal{G}_1 \mathcal{G}_4^*\right]
\nonumber\\&-\hat\Delta_{1Z}\eta_1 {\rm Re}\left[ \mathcal{G}_2 \mathcal{G}_4^*\right]
+\frac{1}{2} \hat s  \hat\Delta_{1Z} {\rm Re}\left[\mathcal{G}_3 \mathcal{G}_4^*\right]\Bigr)+\left(\hat{s}_1\leftrightarrow\hat{s}_2\right)
\end{align}
with $\tilde{\mathcal{G}}_i(s,s_1,s_2)=\mathcal{G}_i(s,s_2,s_1)$ and
\begin{equation}
\eta_1= \hat s\mu_Z-\hat\Delta_{1Z}\hat\Delta _{2 Z},
\end{equation}
\begin{equation}
\eta_2=2 \hat\Delta_{1Z} \hat\Delta _{2 Z}-\hat s \mu _Z,
\end{equation}

\begin{equation}
\eta_3=\left(\hat s_1^2 \hat\Delta _{2 Z}^2-2 \hat s_1 \left(\hat\Delta _{2 Z}-\hat s\right) \mu _Z \hat\Delta _{2 Z}+\mu _Z^2
  \left(\hat\Delta_{2Z} ^2-2\hat s( \hat\Delta_{2Z} +\hat s)\right)\right).
\end{equation}

%\subsection{ $\phi \to Z\gamma\gamma$ ($\phi=h,H$) decay}
 \section{Decay widths of $CP$-even and $CP$-odd scalar bosons}
\label{DecayWidthFormulas}
For completeness, we present the expressions for the most relevant $A\to X$ and $\phi\to X$  ($\phi=h,H$) decays, with $X$ a final multiparticle state. These formulas have been summarized for instance in \cite{Gunion:1989we,Djouadi:2005gi,Djouadi:2005gj}. We use the notation introduced in the Feynman rules shown in Figs. \ref{FeynmanRules1} and \ref{FeynmanRules2}.

\subsection{$CP$-even scalar boson decays}

The tree-level two-body decay width into fermion pairs is

\begin{equation}
\label{phitoff}
\Gamma(\phi\to \bar{f} f)=\frac{f_{\phi \bar{f} f}^2N_c^f m_\phi}{8\pi}\left(1-\tau_f\right)^{3/2},
\end{equation}
with $f_{\phi \bar{f} f}=gm_f  g_{\phi \bar{f} f}/(2m_W)$, where the $g_{\phi \bar{f} f}$ constants are shown in Table \ref{CouplingConstants} for type-II THDM. Also, we use the definition $\tau_a=4m_a^2/m_\phi^2$ and $N_c^f$ stands for the  fermion color number.

The widths of the decays  into a pair of on-shell gauge bosons $V=W\;Z$, when kinematically allowed, are given by
\begin{equation}
\label{phitoVV}
\Gamma(\phi\to VV)=\frac{f_{\phi VV}^2 m_\phi^3}{64n_V\pi m_V^4}\sqrt{1-\tau_V}\left(1-\tau_V+\frac{3}{4}\tau_V^2\right),
\end{equation}
with $n_V=1\; (2)$ for  $V=W\;(Z)$. Here  $f_{\phi WW}=g m_W  g_{\phi WW}$ and $f_{\phi ZZ}=g m_W g_{\phi WW}/c_W^2$, where  again the $g_{\phi VV}$ constants are shown in Table \ref{CouplingConstants} for type-II THDM.

For the present work another  relevant decay is $\phi\to ZA$, whose decay width was already presented in Eq. \eqref{phitoAZ}, which can also be useful to compute the $\phi\to W^\mp H^\pm$ decay when kinematically allowed. On the other hand, we will assume that other tree-level decays such as $\phi\to AA$ and $\phi\to H^-H^+$ are not kinematically allowed and we refrain from presenting the respective decay widths here.

One-loop decays can also be important for Higgs boson phenomenology: while the decay $\phi\to \gamma\gamma$ has a clean signature, the decay $\phi\to gg$ is important for the cross section of Higgs boson production via gluon fusion.  As for the  $\phi\to \gamma\gamma$ decay width, it is given in  Eqs. \eqref{phitoggdecaywidth}-\eqref{phitoggFi}, which can also used for the two-gluon  decay width  by taking the  quark contribution only and making the replacements $\alpha^2\to 2\alpha^2_S$ and $N_c^f Q_f^2\to 1$.

The $\phi\to Z\gamma$ decay has also been largely studied in the literature. The decay width can be written as

\begin{equation}
\label{phitoZgdecaywidth}
\Gamma(\phi\to Z\gamma)=\frac{\alpha^2 m_\phi^3}{512 s_W^2 m_W^2\pi^3}\left(1-\frac{\tau_Z}{4}\right)^3\left|{\mathcal F}^{\phi Z\gamma}\right|^2,
\end{equation}
with ${\mathcal F}^{\phi Z\gamma}={\mathcal F}^{\phi Z\gamma}_f(\tau_f,\xi_f)+{\mathcal F}^{\phi Z\gamma}_W(\tau_f, \xi_W)+{\mathcal F}^{\phi Z\gamma}_{H^\pm}(\tau_{H^\pm},\xi_{H^\pm})$. The contributions of charged fermions, the $W$ gauge boson, and the charged scalar are given by
\begin{equation}
{\mathcal F}^{\phi Z\gamma}_{\chi}(\tau_\chi,\xi_\chi)=
\left\{\begin{array}{lll}\sum_f
\dfrac{2 g_{\phi\bar{f}f}Q_f N_c^f g_V^f}{c_W}\left(I_1(\tau_f,\xi_f)-I_2(\tau_f,\xi_f)\right)&&\chi=f,\\
g_{\phi WW}c_W\left(\left(\left(\frac{2}{\tau_W}+1\right)t_W^2-\frac{2}{\tau_W}-5\right) I_1(\tau_W,\xi_W)+4
\left(3-t_W^2\right)I_2(\tau_W,\xi_W)\right)&&\chi=W,\\
\dfrac{2c_W m_W g_{\phi H^-H^+}}{m_{H^\pm}^2}I_1(\tau_{H^\pm},\xi_{H^\pm})&&\chi=H^\pm,
\end{array}\right.
\end{equation}
where we introduced the  definition $\xi_i=4m_i^2/m_Z^2$.

\subsection{$CP$-odd scalar boson decays}
The decay of a $CP$-odd scalar boson $A$ into a  pair of fermions of distinct flavor is given by
\begin{equation}
\label{Atoff}
\Gamma(A\to \bar{f}f)=\frac{f_{A\bar{f}f}^2N_c^f m_A}{8\pi} \sqrt{1-\tau_f},
\end{equation}
where  now we use the definition $\tau_a=4m_a^2/m_A^2$.

There are no decays into pairs of electroweak gauge bosons at the tree-level, but the $A\to \phi Z$ ($\phi=h,H$) decay can be kinematically allowed. Its decay width is given in Eq. \eqref{phitoAZ} and a similar expression with the corresponding replacements is obeyed by the $A\to W^\pm H^\mp$ decay if kinematically allowed.

As far as one-loop decays are concerned, the two-photon decay proceeds via charged fermion loops  and  its decay width is presented in Eqs. \eqref{phitoggdecaywidth} and \eqref{AtoggFiTot},  whereas the two-gluon decay width can be obtained from these equations by summing over quarks only and making the additional replacements $\alpha^2 \to 2\alpha^2_S$ and $N_c^f Q_f^2\to 1$.

 The $A\to Z\gamma$ decay also receives contribution from charged fermions only and its decay width is given by Eq. \eqref{phitoZgdecaywidth}, with $\phi\to A$ and
\begin{equation}
\mathcal{F}^{AZ\gamma}=\mathcal{F}^{AZ\gamma}_f(\tau_f,\xi_f)=\sum_f \frac{2 g_{A\bar{f}f}Q_f N_c^f g_V^f}{c_W}I_2(\tau_f,\xi_f).
\end{equation}
\subsection{QCD radiative corrrections for the decays $\phi\to \bar{q}q$}
For light quarks, the running mass $\bar m_q$ at the scale $m_\phi$ must be used in Eqs \eqref{phitoff} and \eqref{Atoff} to take into account the next-to-leading order  QCD  corrections. As for higher order QCD corrections, they are important and must be also included. They are summarized in \cite{Djouadi:2005gj} and we include them here for completeness. For light quarks we have
\begin{equation}
\label{QCD}
\Gamma({\phi\to \bar{q}q})=\frac{3g^2 g_{\phi \bar{q} q}^2 \bar{m}_q m_\phi}{32\pi m_W^2}\left(1-\tau_q\right)^{p/2}
\left(1+\Delta_{qq}+\Delta_\phi^2\right),
\end{equation}
where $p=1$ (3) for $CP$-even ($CP$-odd) scalar boson and the running quark mass $\bar m_{ q}$ is defined at the scale $m_\phi$. As for $\Delta_{qq}$, it is the same for both $CP$-even and $CP$-odd scalar bosons for $m_\phi\gg m_q$. In the $\bar{\rm {MS}}$ renormalization scheme it is given by
\begin{equation}
\label{Delta_qq}
\Delta_{qq}=5.67\frac{\bar \alpha_s}{\pi} +(35.94-1.36N_f ) \frac{\bar \alpha_s^2}{\pi^2} +\ldots,
\end{equation}
where $N_f$ is the number of flavors of light quarks and  $\bar \alpha_s$ is the strong coupling constant defined at $m_\phi$ scale. As for $\Delta_\phi$, it differs for $CP$-even or $CP$-odd scalar bosons and it is given at order $\bar\alpha_s^2$ as
\begin{equation}
\label{Delta_phi}
\Delta_ {\phi} =
\frac{\bar\alpha_s}{\pi^2}
\left\{\begin{array}{lcr}
\left(1.57-\frac{2}{3}\log\left(\frac{m_\phi^2}{m_t^2}\right)+\frac{1}{9}\log^2\left(\frac{\bar m_q^2}{m_\phi^2}\right)\right)&&\phi=h,H,\\
\left(3.83-\log\left(\frac{m_\phi^2}{m_t^2}\right)+\frac{1}{6}\log^2\left(\frac{\bar m_q^2}{m_\phi^2}\right)\right)&&\phi=A.
 \end{array}\right..
\end{equation}
For the top quark, the leading order QCD corrections are give by \cite{Djouadi:2005gj}

\begin{equation}
\label{topQCD}
\Gamma({\phi\to \bar{t}t})=\frac{3g^2 g_{\phi \bar{t} t}^2 m_t m_\phi}{32\pi m_W^2}\left(1-\tau_t\right)^{p/2}
\left(1+\frac{4}{3}\frac{\alpha_s}{\phi}\Delta^t_\phi(\beta_t)\right),
\end{equation}
with $\beta_t=1-\tau_t$, whereas $\Delta^t_\phi(\beta)$ is given in Ref. \cite{Djouadi:2005gj}. However, these corrections are small compared to the case of the $b$ and $c$ quarks.

\bibliography{biblio}
\end{document}